\documentclass[lineno]{jfm_arxiv}

\usepackage{graphicx}
\usepackage{xcolor,soul}
\usepackage{mathtools}
\usepackage{newtxtext}
\usepackage{newtxmath}
\usepackage{natbib}
\usepackage{hyperref}
\usepackage{placeins}
\usepackage{cases}
\usepackage{overpic}
\usepackage{siunitx}

\hypersetup{
    colorlinks = true,
    urlcolor   = blue,
    citecolor  = black,
}

\newcommand{\RomanNumeralCaps}[1]
\title{Aerosol deposition in mucus-lined ciliated airways}

\author{Swarnaditya Hazra\aff{1}
 \and Jason R. Picardo\aff{1} \corresp{\email{picardo@iitb.ac.in}}}

\affiliation{\aff{1}Department of Chemical Engineering, Indian Institute of Technology Bombay, Mumbai 400076, India}

\begin{document}

\maketitle

\begin{abstract}
  We study the transport and deposition of inhaled aerosols in a mid-generation, mucus-lined lung airway, with the aim of understanding if and how airborne particles can avoid the mucus and deposit on the airway wall---an outcome that is harmful in case of allergens and pathogens
  %, like dust and virus-bearing droplets, 
  but beneficial in case of aerosolized drugs. %Assuming a Newtonian mucus rheology and focusing on mid-generation airways with laminar airflow, 
  We adopt the weighted-residual integral boundary-layer model of Dietze and Ruyer-Quil (\textit{J. Fluid Mech.}, vol. 762, 2015, pp. 68–109) to describe the dynamics of the mucus-air interface, as well as the flow in both phases. The transport of mucus induced by wall-attached cilia is also considered, via a coarse-grained boundary condition at the base of the mucus.
  %which mimics the metachronal ciliary motion. 
  We show that the capillary-driven Rayleigh-Plateau instability plays an important role in particle deposition by drawing the mucus into large annular humps and leaving substantial areas of the wall exposed to particles. We find, counter-intuitively, that these mucus-depleted zones enlarge on increasing the mucus volume fraction. %We show that this relationship can be explained by contrasting the variation of the wavelength of the fastest-growing instability mode with that of the width of equilibrium-shaped humps (unduloids). 
   Our simulations are eased by the fact that the effects of cilia and air turn out to be rather simple: the long-term interface profile is slowly translated by cilia and is unaffected by the laminar airflow. The streamlines of the airflow, though, are strongly modified by the non-uniform mucus film, and this has important implications for aerosol entrapment. Particles spanning a range of sizes (0.1 to 50 microns) are modelled using the Maxey-Riley equation, augmented with Brownian forces. We find a non-monotonic dependence of deposition on size. Small particles diffuse across streamlines due to Brownian motion, while large particles are thrown off streamlines by inertial forces---particularly when air flows past mucus humps. Intermediate-sized particles are tracer-like and deposit the least. 
  Remarkably, increasing the mucus volume need not increase entrapment: the effect depends on particle-size, because more mucus produces not only deeper humps that intercept inertial particles, but also larger depleted-zones that enable diffusive particles to deposit on the wall. 
  % Increasing the mucus volume produces counter-acting effects: deeper humps that could intercept more particles, and larger depleted-zones that expose more of the wall.  
  %Remarkably, increasing the mucus volume reduces particle entrapment, owing to the widening of depleted-zones which overcomes the effect of deeper humps, and, remarkably, increases deposition on the wall. 
  %On increasing the mucus volume, wider depletion-zones The film profile with increasing volume Increasing the mucus volume fraction deepens the humps but widens the depletion zones. increases the size of 
  %small particles undergo significant Brownian motion and diffuse to the wall, while large particles experience inertial centrifugation across curved streamlines, which occur as the air flows past mucus humps. These humps deepen as the mucus volume is increased and they protrude further into the channel and intercept more particles; however, the depleted zones also widen, exposing larger regions of the wall to particles. We find that the 
  % The latter wins out and, remarkably, more mucus reduces particle entrapment and increases wall-deposition. 
\end{abstract}


\section{Introduction}

The air we inhale brings along a variety of harmful aerosol particles---allergens such as dust and pollen, pollutants like soot, and droplets laden with pathogens---which if deposited on the walls of the airways can cause severe respiratory illnesses \citep{partharm_Beelen}. Furthermore, particles that reach the terminal airways and alveoli can pass into the bloodstream and harm other organs, including the brain and heart \citep{partharm_Fu,partharm_Weichenthal}. The lung's primary defence against airborne particles is provided by mucus which lines the walls of lung airways \citep{Boucher2002,Boucher2006}. The viscous mucus lies atop a sublayer of watery, periciliary liquid (PCL), which bathes a carpet of wall-attached cilia. The tips of the cilia penetrate the bottom of the mucus layer and transport it upward and out of the lungs, thereby evacuating particles that deposit on the mucus \citep{Sleigh1988}.

\begin{figure}
 \includegraphics[width=1.0\textwidth]{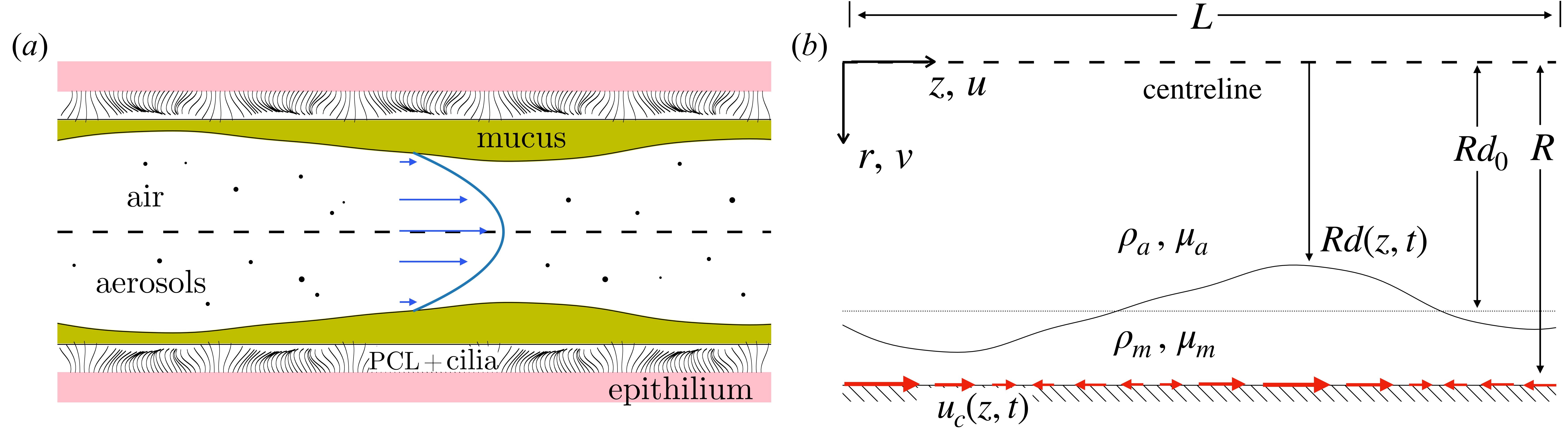}% Here is how to import EPS art
\caption{\label{fig:schematic} (\textit{a}) Illustration of a middle generation, mucus-lined, ciliated airway with inhaled aerosols being transported by the respiratory airflow. (\textit{b}) Schematic of the simplified axisymmetric airway, corresponding to the mathematical model of \S 2. The subscripts $a$ and $m$ denote the air and mucus phases, respectively, and $u_c$ is the spatio-temporally periodic, metachronal velocity (red arrows) imposed by the cilia on the base of the mucus film. }
% Note that $d$ and $d_0$ should be multiplied by $R$ to obtain the dimensional interface position and its uniform initial value.}
\end{figure}

The lungs contain a hierarchy of 24 generations of branching tubular airways, which may be divided into three broad categories: the upper, middle, and terminal airways, corresponding to generations 0 to 9, 10 to 16, and 17 to 23, respectively \citep{Kleinstreuer2010,tsuda2013particle}.
%the upper (from generation 0 to 9), middle (generation 10 to 16), and terminal airways (beyond generation 17) [ref]. 
From the perspective of pulmonary fluid mechanics, these categories are distinguished by the airflow regime (laminar or turbulent), the distribution and composition of the surface liquid, the compliance of the wall, and the presence or absence of cilia. In this work, we focus on a single segment of the middle airways---mucus-bearing, relatively rigid, and ciliated---and study the mucus-entrapment or wall-deposition of airborne particles (see figure~\ref{fig:schematic}\textit{a}). 

In the middle airways, the airflow is laminar with a Reynolds number between 1 to 30, in contrast to the large upper airways where the airflow is turbulent or transitional ~\citep{Kleinstreuer2010,tsuda2013particle}. Though laminar, the airflow in middle airways can be rather complex, owing to the presence of the mucus film, which occupies a substantial portion of the airway. Here, the mucus volume fraction is typically about $10\%$ and can increase further under diseased conditions \citep{Levy2014}. (Mucus is absent from the terminal airways and its volume fraction is relatively small in the upper airways \citep{Sleigh1988}) A key aspect of this two-phase mucus-air flow is the annular interface, which, being endowed with interfacial tension, is susceptible to the Rayleigh-Plateau instability \citep{narayanan}. The mucus film, therefore, is spontaneously driven toward a strongly nonuniform distribution: at low volume fractions the film collects into large annular humps separated by depleted zones \citep{lister2006capillary}, while at higher volume fractions the film can form liquid bridges that block the airway \citep{everett1972model}. The Rayleigh-Plateau instability also produces strong transverse pressure-gradients that cause soft-walled terminal airways to collapse~\citep{heil2008mechanics}; such collapse does not occur in the middle airways thanks to its relatively rigid walls.  

This physical picture of a typical middle airway is completed by cilia, which are present throughout the upper and middle airways~\citep{Levy2014}. Immersed in the low-viscosity PCL, the cilia move synchronously as a metachronal wave with asymmetric forward and backward strokes---the tips of the cilia reach upward and penetrate the bottom of the mucus layer only during the forward stroke~\citep{Sleigh1988}. Under healthy conditions, the stratified arrangement of mucus atop PCL is maintained by a network of crosslinked polymers, within the PCL, that prevents the entry of large molecular-weight mucins~\citep{button2012}. Thus, as a first approximation, it is reasonable to reduce the airway surface liquid to just a single film of mucus with a non-deforming base \citep{romano2022effect}, at which it experiences ciliary forces. The surface of this film is free to deform in response to the action of interfacial tension and airflow (see figure~\ref{fig:schematic}\textit{b}).

%away from a uniform distribution; if the mucus volume fraction is not too large, the airway can remain open but with a strongly non-uniform mucus distribution consisting of bulging unduloids (annular humps) interspersed with mucus-depleted zones. If the volume fraction of mucus is too large, the film forms a liquid bridge and occludes the airway by capillary pressure-gradients to  

Clearly, the middle airways present a challenging fluid mechanical problem, one that spans a wide range of length and time scales (see table~\ref{table:properties}). The flow and particle transport therein play an important role in respiratory diseases, because (i) if harmful particles pass through untrapped, then they will enter deep into the lungs where there is no mucus barrier, and (ii) oversecretion of mucus or airway constriction, triggered by the deposition of inhaled allergens~\citep{allergen2015}, can produce mucus plugs that obstruct airflow. Plugs form more easily in the smaller middle-airways (as compared to the larger upper-airways) because of the relatively large volume fraction of mucus and the relatively weak airflow, which struggles to expel the mucus plugs. Such plugs are commonly observed in cases of rapid-onset, fatal asthma~\citep{hays2003role,rogers2004airway}. The most effective treatment for asthma, and related diseases like chronic bronchitis, involves inhalation of aerosolized drugs such as bronchodilators that relax the muscles of the airways~\citep{rogers2004airway,bronchodilator}. In this flipped scenario, the mucus film limits the efficacy of inhalation therapy~\citep{mucus_drugs}, which requires that drug particles avoid mucus-entrapment and deposit on exposed sections of the airway wall.

\begin{table}
\begin{center}
\phantom{~}\noindent
\begin{tabular}{>{\raggedright\arraybackslash}p{0.3\textwidth}@{\hskip 30pt}>{\raggedright\arraybackslash}p{0.2\textwidth}>{\raggedleft\arraybackslash}p{0.25\textwidth}}
Property & Typical value  & Reference\\ [12pt]
% \hline
 % \multicolumn{3}{l}{length scales  (\si{mm})}\\[6pt]
  \textit{length scales} & \si{mm}  & \\ [2pt]
% \hline

airway diameter&  $1$ &\citet{pedley1977pulmonary} \\
mucus layer thickness& $0.01$  & \citet{Sleigh1988} \\
PCL layer thickness& $0.005$  &  \citet{Sleigh1988} \\
cilia height& $0.005 $  & \citet{Sleigh1988} \\
% \citet{oldenburg2012monitoring} \\
metachronal-wavelength & $0.03$ & \citet{fulford1986muco} \\
airway length & $5 - 10$  & \citet{Sleigh1988} \\ 
wavelength of RP instability & 3 & \\
distance travelled by a tracer per breathing cycle & $3$  & \\ [22pt]

% \hline
 % \multicolumn{3}{l}{velocity scales (\si{mm.s^{-1}})}\\
  \textit{velocity scales} & \si{mm.s^{-1}}  & \\ [2pt]
% \hline

celerity of ciliary wave& $0.3$  &  \\
mucociliary transport velocity & $0.05$  & \citet{vasquez2016modeling}; \citet{oldenburg2012monitoring} \\
% Mucus velocity(hump formation)& $5 - 10$  & \citet{Sleigh1988} \\
air velocity & $10$  & \citet{tsuda2013particle} \\ [12pt]
% \hline
 % \multicolumn{3}{l}{time scales  (\si{s})}\\
  \textit{time scales} & \si{s} & \\ [2pt]
% \hline

cilia-beat time period & $0.1$  & \citet{chatelin2016parametric} \\
breathing time period& $1$  & \citet{pedley1977pulmonary} \\
mucociliary transport time (across airway length)& $100$  &  \\ [12pt]
% \hline
 % \multicolumn{3}{l}{non-dimensional numbers}\\
 non-dimensional numbers & $[\cdot]$ & \\ [2pt]
% \hline

air Reynolds number & $1 - 30$  & \citet{pedley1977pulmonary} \\
 mucus Reynolds number& 
 % $2.3\times 10^ {-3}$
 $10^ {-3}$  &  \\
air-mucus viscosity ratio & $10^{-4} - 10^{-3}$  & \citet{erken2022capillary} \\
% air-mucus viscosity ratio & $0.0002 - 0.006$  & \citet{erken2022capillary} \\ & & \citet{Dietze2015} \\
% air-mucus viscosity ratio & $0.0002 - 0.006$  & \citet{erken2022capillary} \hspace{60pt} \citet{Dietze2015} \\
air-mucus density ratio & $10^{-3}$  & \citet{erken2022capillary} \\
% \hline
\end{tabular}
\caption{Typical magnitudes of various length, velocity, and time scales in a mid-generation airway, highlighting the multiscale nature of the associated transport phenomena.}
\label{table:properties}
\end{center}
\end{table}

Despite its importance, flow in the middle airways has not received much attention, possibly because it involves the simultaneous interaction of air, mucus, and cilia~\citep{Levy2014}. Most previous studies have focused either on mucus transport by cilia (ignoring air and capillary dynamics of the interface) or  air-mucus dynamics (ignoring cilia). The former combination is acceptable for the upper airways where the mucus volume fraction is too low to obstruct the airflow, while the latter combination is certainly relevant to terminal airways which are devoid of cilia. The middle airways, however, require all three factors to be considered together, which is what we do here, albeit with simplifying assumptions. Knowledge of the air-mucus flow allows us to track the motion of air-borne particles and study their deposition on the cilia-transported, non-uniform mucus film. We thus address the following questions: (i) Does the Rayleigh-Plateau instability, which yields a non-uniform mucus distribution of humps and depleted zones, aid mucus-entrapment or facilitate wall-deposition? (ii) How do changes in the mucus volume-fraction affect particle deposition? (iii) What is the effect of the particle's size on its entrapment? 

Before proceeding, though, it is useful to briefly summarize our current understanding of the three physical processes that are central to the lung's particle-trapping-evacuation system: (a) capillary-driven film dynamics, (b) mucociliary transport, and (c) particle deposition.

An annular film on the inner wall of a cylindrical tube is driven by the Rayleigh-Plateau instability to one of two morphologies. At low volume fractions, the film accumulates into unduloids, i.e. equilibrium-shaped annular humps or collars, which are separated by depleted zones that are devoid of liquid. For volume fractions beyond a critical value, unduloid solutions cease to exist and instead the film forms liquid bridges \citep{everett1972model}. Not only do these bridges block the airway, their formation is associated with strong recoil forces which can damage the cells of the airway wall~\citep{Romano2019viscous,erken2022capillary}. Several studies have therefore analysed liquid-bridge formation, considering the effects of mucus rheology~\citep{Romano2021viscoelastic,Romano2023viscoplastic} and surfactants~\citep{romano2022effect}. If the wall is soft, as in the terminal airways, closure can occur even at lower volume fractions, because the capillary-induced reduction of pressure inside liquid humps causes the wall to collapse~\citep{heil2008mechanics}. Here, we are interested in non-collapsing, rigid-walled airways with mucus fractions that are low enough to avoid closure. Even so, the Rayleigh-Plateau instability remains important because it controls the sizes of mucus humps and depleted zones. On the one hand, humps protrude into the airway and may intercept airborne particles; on the other hand, depleted zones leave the wall exposed to particle deposition. We shall see that the net outcome of these competing effects depends on the size of the particles. 

Does airflow alter the distribution of mucus? Because mucus has a much higher density and viscosity (table~\ref{table:properties}), the influence of airflow on the morphology of the film is restricted to the upper airways, where the airflow is turbulent. In the extreme case of a cough or sneeze, rapidly flowing air can rip off droplets of mucus from the film \citep{mittal2020flow,morawska2022physics}. In the middle airways, though, the airflow is laminar and too weak, under normal breathing conditions, to appreciably influence the morphology of the mucus film. (Our simulations, in \S~\ref{sec:air}, confirm this assessment.) So, we include respiratory airflow in the present work not for its influence on the mucus film but to track the motion of airborne particles through the airway.

Mucociliary transport depends on the asymmetric and synchronous beating of wall-attached cilia. Each cilium executes a periodic three-dimensional motion with a quick upright forward stroke, and a slow bent-over backward stroke~\citep{Sleigh1988,Blake2001bookch}. This motion is produced by internal active forces exerted by molecular motors~\citep{gilpin2020ciliarev} and is influenced by the flexibility of the cilium, as well as hydrodynamic interactions between segments of the cilium with each other and with the wall. Inter-cilia hydrodynamic interactions play an important role in establishing a phase-shifted synchronization that produces a metachronal wave which sweeps across the carpet of cilia~\citep{mitran2007synchronization,biciliaHI2018,ciliaHI2020}; the details of this complex fluid-structure interaction problem continue to be investigated~\citep{Chakrabarti2019HI,Chakrabarti2022sync}. Studies of mucociliary transport have typically imposed this synchronized pattern of motion onto model cilium and then calculated the resulting mucus flow~\citep{smith2008modelling}. Early work, pioneered by Blake, treated the cilium using a string of Stokeslets~\citep{Smith2007Stokeslet,Smith2009stokesletelastic,smith2018Blaketribute}, while more recent work has used techniques like the immersed boundary method~\citep{Sedaghat2016lbm}. Models have also progressed from considering just the mucus phase to including both the PCL and mucus layers~\citep{Jayathilake2020,quek2018three}. %The viscoelastic Non-newtonian rheology has also been accounted for [ref]. 
The computational expense of discrete-cilia models has motivated the formulation of parsimonious coarse-grained models in which explicit cilia are replaced by a force distribution \citep{smith2006viscoelastic,smith2007model} or a boundary condition at the base of the mucus~\citep{vasquez2016modeling,Bottier2017exp}. Such a simple ciliary forcing has been used to compute large-scale flow patterns \citep{vasquez2016modeling} and to gain analytical insight into the role of viscoelasticity \citep{Dietze2023mucociliary}. We shall adopt the boundary-condition representation of cilia here as well.

Mucus is certainly non-Newtonian with viscoelastic properties~\citep{Levy2014}. Its viscoelasticity plays a crucial role during rapid flows, such as the capillary-driven flow associated with airway closure and reopening~\citep{Romano2021viscoelastic}. Here, we focus primarily on open airways with relatively low mucus volumes; the corresponding slow thin-film flow is not strongly affected by viscoelasticity \citep{halpern2010effect}. 
%thin films is slow and viscoelascity in which the thin film flow is too slow for cases in which the airway stay open;  mucus volume is low enough for the airway to remain open for which closure does not occur; the   as well as the formation of annular humps~\citep{halpern2010effect}. However, once capillary pressure-gradients have relaxed, having produced quasi-static humps, the subsequent slow flow will not depend as strongly on viscoelasticity. 
Even with cilia-driven oscillatory forcing, \citet{Dietze2023mucociliary} have shown that the effects of viscoelasticity are marginal under healthy conditions (with diseased mucus there is a reduction in mucociliary transport). 
%Our present study is focused on the long-term profile of the mucus film and its impact on particle transport in healthy and open airways; 
So, for simplicity, we henceforth neglect viscoelasticity.

%A rare study combining both mucociliary transport and film dynamics is that of Dietze [ref], who account for ciliary forcing by applying a traction boundary condition to the base of the mucus film. The film is seen to take the the form of a periodic sequence of travelling mucus humps. Dietze [ref] also show that the viscoelasticity of mucus affects the results quantitatively, increasing the humps' celerity, but not qualitatively. Indeed, viscoelasticity is known to play a crucial role during rapid flows, as occurs during the initial formation of humps [ref] or liquid-bridges [ref]. However, once capillary pressure-gradients have relaxed, having produced quasi-static humps, the subsequent slow cilia-induced flow will not depend as strongly on viscoelasticity. It is this long-term state of the mucus film that is relevant for our study of particle transport, and so for simplicity we henceforth neglect viscoelasticity.

Particle deposition in the lungs has been measured experimentally and found to depend strongly and non-monotonically on the size of the particles, with minimal deposition at intermediate sizes~\citep{Heyder1986totaldeposition,morawska2022physics}. The physical mechanism that drives deposition varies with particle size \citep{guha2008transport}, which ranges from about 5 nm to 50 $\mu$m, and with the generation of the airway~\citep{tsuda2013particle}. The smallest particles, which are strongly affected by Brownian forces, diffusive across air streamlines and deposit on the wall or mucus. This diffusive deposition of tiny particles is relevant throughout the lungs. Intermediate-sized particles behave like tracers, following air streamlines, and so deposit the least~\citep{tsuda2013particle}.

The larger particles are unaffected by Brownian forces but deviate from streamlines due to their inertia. Whenever air streamlines are curved, inertial particles will experience cross-streamline centrifugal forces that could thrust them towards the wall \citep{guha2008transport}. Indeed, inertial deposition is known to be important in the upper airways, where the turbulent flow is strongly vortical, and at airway bifurcations, where secondary circulations are present~\citep{Kleinstreuer2010,tsuda2013particle}. (Centrifugal ejection of inertial particles from vortices~\citep{Maxey1987settling} has been well-studied in the context of turbulent suspensions, such as droplet-laden warm clouds, for its role in accelerating collisions~\citep{Grabowski2013,Ravichandran2020}). In this study, we will show that inertial deposition is relevant even in straight middle-airways, where the airflow is laminar, because of the presence of mucus humps which force the air streamlines to curve around them. 

In the terminal airways, the airflow is too weak for inertial forces to play a role. Instead, gravity aids in the deposition of large particles. In the higher generations, though, gravity is unimportant, being overwhelmed by inertia and air-drag, in case of large particles, or Brownian forces, in case of small particles~\citep{tsuda2013particle}. 

Simulations of particle deposition in the lungs are challenging because of complex airway geometries and air flow fields. Nevertheless, computational studies have been carried out for the nasal passage \citep{nasal2009,nasal2016,nasal2020}, the trachea and upper conducting airways \citep{trachea1995,trachea2007,upper2015}, and airway bifurcations \citep{branch2004,Kleinstreuer2010,Panchagnula-branching-24}. In these cases, the flow is typically turbulent or transitional. At the other end, transport in the terminal alveoli-bearing ducts of the pulmonary acinus has also received much attention \citep{tsuda2013particle}; in fact, realistic simulations have been performed in computational geometries constructed by imaging rat lungs \citep{tsuda2008xray}. In the alveoli, where the airflow is in the creeping regime, particle transport is facilitated by chaotic advection \citep{Tsuda2002chaotic,tsuda2011chaotic,Panchagnula-chaotic}. 

The question of how the mucus film affects particle deposition has largely been overlooked, with the exception of a few studies \citep{ahmadi2023mucus-nasal}, including the important work by Kim and coworkers: In a pair of studies, one in vitro \citep{kim1985in-vitro} and the other in vivo \citep{kim1985in-vivo}, they show that the presence of the mucus film increases the rate of particle deposition. These studies consider an upper-generation bronchus and find that the relatively rapid airflow produces waves on the mucus film. In the middle airways, with which we are here concerned, the airflow is nearly a hundred times slower under normal breathing conditions \citep{tsuda2013particle}; so we do not expect waves to form on the film. Nonetheless, we expect the mucus film, with its humps and depleted zones, to impact particle deposition by altering the path of airflow and the surface area for deposition. 
% Whether these film-induced effects result in more mucus entrapment (and evacuation) or more exposed-wall deposition (and possible infection or drug-delivery) is a question we shall answer in this paper.

We address this multiscale problem using long-wave reduced-order modelling techniques, along with various simplifying assumptions. The airway is modelled as a straight, rigid, cylindrical tube. The mucus layer is treated as a Newtonian viscous liquid film lining the inner cylindrical wall. The ciliary forcing to the base of the mucus is modelled as a time-dependent velocity boundary condition at the wall, designed to mimic the asymmetric metachronal wave of the cilia. No-penetration is also imposed at the wall considering that, in healthy airways, mucus is prevented from entering the PCL sub-layer by a network of cross-linked polymers \citep{button2012}. Thus, like many prior models of the airway surface liquid \citep[\textit{e.g.},][]{Romano2019viscous,Romano2023viscoplastic}, we subsume the PCL-cilia layer into the wall and solve only for the dynamics of the mucus layer. The coupled flow of air and mucus---and the dynamics of the interface---is described by the two-phase thin film equations obtained from the weighted-residual integral boundary layer (WRIBL) method of \citep{Dietze2015}. This method improves upon traditional long-wave lubrication theory and consistently accounts for (i) inertia up to $\Rey \sim 10$ \citep{Dietze2013}, as occurs in the air, (ii) longitudinal viscous stresses which are relevant in thinning necks of the viscous mucus film \citep{Dietze2015}, and (ii) nonlinear interfacial curvature which is crucial for realising airway closure. The WRIBL model has been extensively validated against direct numerical simulations, including for core-annular air-mucus flow \citep{Dietze2013,Dietze2015}. We further assume an axisymmetric flow field that is periodic in the longitudinal direction. 

Particles are evolved assuming one-way coupling with air and ignoring inter-particle collisions (dilute limit). Within the point-particle approximation, their motion is described by the simplified Maxey-Riley equation for tiny heavy spherical particles~\citep{maxey1983,ravichandran-rev2017}, augmented by Brownian noise. The resulting Langevin equation can be used for the full range of inhaled aerosols, with the relative strength of Brownian and inertial forces being determined by the particle size~\citep{tsuda2013particle}. 

The mathematical model is presented in \S 2; certain simplifying limits of the model are also discussed, along with the corresponding numerical solution procedures. Section 3 focuses on just the air-mucus flow without particles. 
We show that interfacial deformation is driven by capillary forces and is unaffected by air (whose viscosity and density are much smaller than mucus) and ciliary transport; the latter only produce a slow lateral transport of the deformed film. These findings allow us to simplify the model and thereby ease subsequent computations. 
% The physical parameters suggest that interfacial deformation is driven by capillary forces and is unaffected by either air (its viscosity and density are much smaller than mucus) or cilia (the ciliary velocity is much smaller than that of the capillary-driven mucus flow during hump formation), though the latter should produce a slow lateral transport of the deformed film. We confirm this expectation and show that the mucus film forms a sequence of humps and depleted zones which are simply translated with the mean cilia velocity. These findings allow us to simplify the model and thereby ease subsequent computations. 

Next, in \S 4, we take up the important question of how the extent of depleted zones---which expose the airway wall to particles---varies with the mucus volume fraction. The variation, measured from a number of randomly initialized simulations, is counter-intuitive: increasing the mucus volume fraction increases the extent of depleted zones. This behaviour is shown to arise from the manner in which equilibrium unduloids change shape as their volume increases.

Particle deposition is examined in \S 5, considering two scenarios, distinguished by the extent of particle turnover after each breath. The typical distance travelled by a particle in a breathing cycle is less than the length of an airway (table~\ref{table:properties}); so particles could remain within the same airway for many breathing cycles. However, inhalation and exhalation will bring fresh particles into the airway and remove old ones; mixing with upstream and downstream airways will also contribute to the turnover of particles (\citeauthor{airwaymixing1959} \citeyear{airwaymixing1959}; \citeauthor{Wang2005book} \citeyear{Wang2005book}, chap 5). To aid understanding, we study the two opposing extremes: (i) particles initially introduced into the airway remain within the airway throughout the simulation and no new particles are introduced, \textit{i.e.}, there is no particle turnover; (ii) particles are completely replaced after every breath. Simulations are performed for dozens of breathing cycles (about a minute of breathing). Starting with the first scenario, we characterize the non-monotonic dependence of deposition on particle size and highlight the effect of the mucus volume fraction. The results are explained in terms of the physical mechanisms of Brownian and inertial cross-stream motion, and the distribution of particles in the airway. This understanding carries over to the second scenario, which reveals the effects of respiratory particle-turnover. 

In the concluding \S 6, we summarize the key results and discuss their implications for future work.

\section{Mathematical model}\label{sec:mathmodel}

\subsection{Long-wave equations}\label{sec:goveqn}

We model the airway as a straight cylindrical tube of radius $R$, lined by an annular film of mucus, which is assumed to be a Newtonian fluid with viscosity $\mu_m$ and density $\rho_m$. Air of viscosity $\mu_a$ and density $\rho_a$ flows through the core. The mucus film interacts with air through an interface that is endowed with interfacial tension $\gamma$. We assume axisymmetry and adopt a cylindrical coordinate system (see figure~\ref{fig:schematic}\textit{b}). 
% with scaled coordinates $r$ and $z$ in the radial and axial directions, respectively. 
The axial length scale $\Lambda$ that characterizes axial variations is assumed to be much larger than the radius $R$ so that $\epsilon = R/\Lambda$ is a small parameter. We can then treat the two-phase flow in the thin film limit and obtain simplified long-wave equations. Following \citet{Dietze2015}, we scale the continuity and Navier-Stokes equations and retain terms up to $\mathcal{O}(\epsilon^2)$ to obtain the following non-dimensional, long-wave, boundary-layer equations \citep*{Kalliadasis} for pressure, $p_i$, and the radial and axial velocities, $v_i$ and $u_i$:
\begin{equation}\label{continuity}
\frac{1}{r}{\partial_r}(r v_i) + {\partial_z u_i} = 0,
\end{equation}
\begin{multline}\label{BLE}
 \epsilon Re_i\left(\partial_t u_i + v_i\partial_r u_i+ u_i\partial_z u_i\right)= -\partial_z \left(p_i |_{d}\right) -\epsilon^2 \partial_z \left(\partial_z u_i  |_{d}\right) +\frac{1}{r}\partial_r \left(r\partial_r u_i\right) \\
+2\epsilon^2 \partial_{zz} u_i+ G_{i}sin( \omega t),
\end{multline}
where all variables are non-dimensional, and $\partial_t$ represents the partial derivative with respect to time $t$, and so on for $\partial_r$ and $\partial_z$. The interface between the air and mucus (denoted by subscripts $i = a, \, m$) is located at $r = d(z,t)$. The two terms evaluated at the interface, \textit{i.e.} those involving $p_i |_{d}$ and $u_i  |_{d}$, were obtained by substituting for $\partial_z p_i$ the expression obtained after integrating the $\mathcal{O}(\epsilon^2)$ radial momentum equation from $r$ to $d$. 

The following characteristic scales (decorated by *) have been used for non-dimensionalization: 
\begin{equation}\label{scales}
  r^* = {R},\, z^* ={\Lambda},\, t^* = {R/U},\,
  d^* = {R},\,  u^*_i ={U}, \,  v^*_i =\epsilon {U}, \, p_i^* = \frac{\mu_i U}{\epsilon R},
\end{equation}
where $U$, the common velocity scale for both phases, will be chosen shortly. 

The last term of \eqref{BLE} arises from the imposition of an oscillatory body force (or a background, uniform, axial pressure-gradient) in the air phase ($G_m = 0$), of frequency 
% $f_b =\omega U/2 \pi R$ 
$f_b =\omega /2 \pi $ and dimensional amplitude $A = G_a \mu_a U/R^2$, to mimic respiratory airflow. We now select the velocity scale $U$ based on the airflow, by setting $G_a = 1$ so that $U = A R^2/\mu_a$. Then, the non-dimensional Reynolds numbers, $\Rey_i = \rho_i U R/{\mu_i}$, become $\Rey_a= \rho_a A R^3/ \mu_a^2$ and $\Rey_m = \Rey_a \Pi_{\mu}/\Pi_{\rho}$, where  $\Pi_{\mu} = \mu_a/\mu_m$ and $\Pi_{\rho} = \rho_a/\rho_m$ are the viscosity and density ratios.

% along with the viscosity and density ratios,  $\Pi_{\mu} = \mu_a/\mu_m$ and $\Pi_{\rho} = \rho_a/\rho_m$, account for the physical properties of air and mucus system.
% \begin{equation}
%   z =\frac{z^*}{L},\,  r =\frac{r^*}{R},\, t =\frac{t^*}{R/U_{\gamma}},
%   d =\frac{d^*}{R},\, p_i = \frac{p_i^*\epsilon}{\mu U_{\gamma}/R},\,  u_i = \frac{u^{*}_i}{U_{\gamma}}
% \end{equation}

Next, we list the conditions at the boundaries, starting with the interface, $r = d(z,t)$, where we require the velocities to be continuous,
\begin{align}\label{vel_con}
u_a = u_m , \quad v_a = v_m, \quad at \quad r = d,
\end{align}
and apply the $\mathcal{O}(\epsilon^2)$ balances of tangential and normal stresses:
 \begin{equation}\label{tangential}
   \partial_r u_m-\Pi_\mu \partial_r u_a = \left[2\epsilon^2 \partial_z d\left(\partial_z u_m- \partial_r v_m\right)-\epsilon^2 \partial_z v_m\right] - \Pi_\mu\left[2\epsilon^2 \partial_z d\left(\partial_z u_a-\partial_r v_a\right)-\epsilon^2 \partial_z v_a\right],
 \end{equation}
\begin{equation}\label{normal}
 p_m - \Pi_{\mu} p_a =  - Ca (\kappa) + 2\epsilon^2\left(\partial_r v_m- \partial_r u_m \partial_z d \right)
-2\epsilon^2 \Pi_\mu\left(\partial _r v_a- \partial_r u_a \partial_z d \right),
 \end{equation}
 where $Ca = \gamma/\mu_m U $ is the capillary number and the mean-curvature $\kappa$ to  $\mathcal{O}(\epsilon^2)$ is given by
\begin{equation}\label{curvature}
  \kappa = \frac{1}{d}-\frac{\epsilon^2\left(\partial_z d\right)^2}{2d}-\epsilon^2 \partial_{zz}.
\end{equation}
This nonlinear approximation of the full curvature is essential for capturing the onset of liquid-bridge formation \citep{Dietze2015}, which limits the volume of mucus that can be occupied in an open airway; we will show later that nonlinear curvature also has an important effect on the extent of mucus-depleted zones at the wall. In fact, the primary reason we retain terms up to $\mathcal{O}(\epsilon^2)$ is to consistently include these effects of nonlinear curvature in the model.

The evolution of the interface is governed by the kinematic boundary condition:
\begin{equation}\label{kinematic-gov}
\partial_t d = v_m - u_m\partial_z d.   
\end{equation}

At the centreline, we have the symmetry condition,
\begin{align}\label{symmetry}
 v_a = 0, \quad \partial_r u_a = 0 , \quad at \quad r = 0,
\end{align}
while at the wall, we apply the non-penetration condition and impose a tangential velocity to account for mucociliary transport: 
\begin{align}\label{no_pen_no_slip}
 v_m = 0, \quad u_m = u_c(z,t), \quad at \quad r = 1.
\end{align}
The function $u_c(z,t)$ is a travelling wave, chosen to mimic the asymmetric metachronal wave of the cilia carpet:
\begin{equation}\label{cilia_bc}
   u_c(z,t) = 
     \begin{cases}
       % \text{$a_f\langle\, u_c \rangle \,sin\left(2\pi(z-ct)/\lambda_c\right)$} \quad 2n\pi \le z-ct < 2n\pi+\pi\\ 
       % \text{$a_r\langle\, u_c \rangle \,sin\left(2\pi(z-ct)/\lambda_c\right)$} \quad 2n+ \pi \le z-ct < 2n\pi+ 2\pi\\
        \text{$a_f\langle\, u_c \rangle \,sin\left(2\pi(z+ct)/\lambda_c\right)$} \quad n\lambda_c \le z+ct < n\lambda_c+\lambda_c/2\\ 
       \text{$a_r\langle\, u_c \rangle \,sin\left(2\pi(z+ct)/\lambda_c\right)$} \quad n\lambda_c+\lambda_c/2 \le z+ct < n\lambda_c+\lambda_c\\
     \end{cases},
\end{equation}
for $n = 0,1,2...$. The celerity of the wave is given by $c =  f_c \lambda_c$ where  $\lambda_c$ is the metachronal wave length and $f_c$ is the cilia-beating frequency. The mean cilia velocity, $\langle\, u_c \rangle = \frac{1}{\lambda_c}\int_{0}^{\lambda_c} u_c(z',t)dz'= \frac{1}{f_c^{-1}}\int_{0}^{f_c^{-1}} u_c(z,t')dt'$, is a non-zero constant because the forward stroke is stronger than the reverse stroke,\textit{ i.e.}, the ratio of the constants $a_f/a_r$ is greater than unity. Note that the metachronal wave is antiplectic and propagates in the direction opposite to that of the effective cilia stroke.
% Note that because $u_c$ has the form of a the travelling-wave, $\langle\, u_c \rangle_z = \langle\, u_c \rangle_t = \frac{1}{f_c^{-1}}\int_{0}^{f_c^{-1}} u_c(\zeta)d\zeta$.

This cilia boundary condition is similar to that used by \citet{vasquez2016modeling} to model cilia-driven large-scale swirling flows, observed in an in vitro experiment. However, while \citet{vasquez2016modeling} use a spatially invariant boundary velocity, we have used a travelling wave to better represent the metachronal cilia motion. Our boundary prescription is also closely related to that developed and validated by \citet{Bottier2017model} and used recently by \citet{Dietze2023mucociliary} to understand the influence of viscoelasticity on mucociliary transport; this is a Navier-slip condition that contains a slip length, which when set to zero yields a Dirichlet boundary condition just like ours. We proceed with a zero-slip condition for simplicity. 
As demonstrated below, the cilia-induced velocity is much too slow to alter the shape of the film or the streamlines of the airflow, and so our results regarding particle deposition will not change with the addition of slip or other modifications to the precise form of the cilia boundary condition. 

To ease computations, we shall later explore whether the travelling-wave cilia-velocity can be replaced by just its mean value $\langle\, u_c \rangle$; the resulting Couette boundary condition is
\begin{equation}\label{cilia_bc_const}
  v_m = 0, \quad  u_m = \langle\, u_c \rangle, \quad at \quad r = 1.
\end{equation}

In the axial direction, we consider periodic boundary conditions. Thus, the interface profile $d(z,t)$ evolves 
without changing the total volume of mucus, which remains at its initial value of $(1-d_0^2) L R^2$, where $d_0$ is the dimensionless position of the initially-flat film ($d(z,0) = d_0$) and $L$ is the length of the domain (see figure~\ref{fig:schematic}). 

Equations \eqref{continuity}, \eqref{BLE}, \eqref{vel_con} - \eqref{symmetry}, along with either \eqref{no_pen_no_slip} and \eqref{cilia_bc}, or just  \eqref{cilia_bc_const}, form a closed system of equations. Most of the parameters are set to physiologically relevant values, listed in table~\ref{table:simulation-parameter}. The initial film thickness, $1-d_0$, is varied to study the effect of the mucus volume fraction. 

\begin{table}
\begin{center}
\phantom{~}\noindent
\begingroup
\renewcommand{\arraystretch}{1.2}
\begin{tabular}{>{\raggedright\arraybackslash}p{0.25\textwidth}>{\centering}p{0.1\textwidth}>{\centering}p{0.12\textwidth}>{\raggedright\arraybackslash}p{0.15\textwidth}>{\raggedleft\arraybackslash}p{0.27\textwidth}}
Parameter & Unit & Symbol & Value & Reference\\[10pt]
% \hline
% \multicolumn{4}{c}{Airways}\\[3pt]
% \hline
 
airway radius & mm & $R$ &$ 0.4 $ & \citet{romano2022effect}; \citet{weibel1963morphometry}\\
% \hline
% \multicolumn{4}{c}{Mucus}\\[3pt]
% \hline

mucus viscosity & \si{Pa.s}& $\mu_m$ & $0.01$ & \citet{erken2022capillary}\\
air viscosity & \si{Pa.s}& $\mu_a$ &  $1.8\times 10^{-5}$  & \citet{Dietze2015}\\
mucus density & \si{kg.m^{-3}}  & $\rho_m$& $1000$ & \citet{erken2022capillary}\\
air density & \si{kg.m^{-3}} & $\rho_a$ &$1.2$ & \citet{Dietze2015}\\
interfacial tension & \si{N.m^{-1}}& $\gamma$ & $ 0.05 $ & \citet{erken2022capillary}\\
% \hline
% \multicolumn{4}{c}{Airflow}\\[3pt]
% \hline       
airflow-driving force amplitude &\si{N.m^{-3}}  &$A$& $18$ & chosen to match typical flow rates: \citet{pedley1977pulmonary}\\
velocity scale & \si{m.s^{-1}} & $U$ & $A R^2/\mu_a$ & \\
respiratory frequency & Hz  &$f_b \,U/R$& $1$ & \citet{pedley1977pulmonary}\\
% \hline
% \multicolumn{4}{c}{Cilia}\\[3pt]
% \hline
mean/net cilia velocity & \si{\mu m.s^{-1}} & $\langle u_c \rangle U$  & $40$ & chosen to match typical transport rates:\citep{Dietze2023mucociliary}\\
cilia forward to reverse stroke ratio & $[\cdot]$   &$a_f/a_r$& $2 $ & \citet{vasquez2016modeling}\\
% cilia reverse stroke &   $\si{mm.s^{-1}}$ &$a_r$& $1.26 \times10^{-2} $ & \citet{vasquez2016modeling}\\
cilia-beat frequency & Hz &$f_c \,U/R$& $10 $ & \citet{oldenburg2012monitoring}\\
metachronal wavelength & mm &$\lambda_c R$& $40\times 10^{-3} $ & \citet{blake1975movement}\\
% \hline
% \multicolumn{4}{c}{Non-dimensional parameter}\\[3pt]
% \hline
Reynolds number of air &  $[\cdot]$  &$\Rey_{a}$& $4$ & \\
Reynolds number of mucus &  $[\cdot]$ &$\Rey_{m}$& $6.1$ & \\
Capillary {number} &  $[\cdot]$ &$Ca$& $32$ & \\
viscosity ratio &  $[\cdot]$ &$\Pi_{\mu}$& $1.8\times 10^{-3} $ & \\
density ratio &  $[\cdot]$  &$\Pi_{\rho}$& $1\times 10^{-3} $ & \\
initial film thickness &  mm  &$(1-d_0) R$& $0.008$ - $0.045$ & \\
fastest-growing Rayleigh-Plateau wavelength & mm  & $\Lambda_{RP} $ & $3.1$ - $3.4$ & 
% $2\pi 2^{1/2} d_0$: 
\citet{Rayleigh1892wavelength}\\
computational domain length& mm  &$ L$& $ \Lambda_{RP}$  & \\
% Volume  &$V$& $1.70 $ & -\\[2pt]
\end{tabular}
\endgroup
 \caption{Values of the parameters in the flow model, corresponding to the results in the main text. Some key figures are also produced for a second set of mucus-air properties in the \href{https://bighome.iitb.ac.in/index.php/s/6p6PHMqSKYHD6sx}{ supplementary material}. 
% \textcolor{red}{To illustrate the effect of ciliary transport on closure, we also consider a much slower mean cilia speed of $4$ \si{\mu m.s^{-1}}, instead of  $40$ \si{\mu m.s^{-1}}, in \S~\ref{sec:depleted}}.
}
\label{table:simulation-parameter}
\end{center}
\end{table}

\subsection{Weighted-residual approach}
We now average across the radial direction, using the WRIBL method, to obtain evolutions equations for the interface profile $d(z,t)$ and the flow rates of air and mucus, $Q_i(z,t)$. The derivation is outlined, in brief, in this section; for details, the reader is referred to \citet{Dietze2015}, whom we follow closely. The implementation of this derivation using computer algebra is discussed in \citet{HazraWRIBL}.
% In WRIBL method the velocity field is expanded into two parts. A leading order term and a correction term.  When we put that back into the BLEs \ref{BLE}, then the correction term only survived in the cross stream viscous diffusion term. That can be eliminated by choosing suitable weight functions\ref{BLE} of the respective phases. Then has been multiplied them with respective BLE's.  

% Give leading and correction decomposition (purely epsilon), and outline procedure. Maybe say that leading order component will be used for streamlines and thus is given below. 

\subsubsection{Velocity decomposition}
The velocity field is decomposed into a leading-order contribution $\hat{u}_i$ and a correction $u'_i$  that is $\mathcal{O}(\epsilon)$. The leading term accounts for the local balance between the $\mathcal{O}(1)$ radial viscous diffusion and axial pressure gradients, while yielding the exact flow rate $2 \pi Q_i$. Thus,  $\hat{u}_i$ is a self-similar radial profile that is parameterized by the cilia velocity, the flow rate, and the film thickness:
\begin{equation}\label{vel_decomp}
    u_i(r,z,t) = \hat{u}_i(r;u_c,d,Q_i) + u'_i(r,z,t),
\end{equation}
with 
\begin{equation}\label{u_hat}
  \quad \frac{1}{r}\partial_r \left(r\partial_r \hat{u}_m\right) = A_m, \quad \frac{1}{r}\partial_r \left(r\partial_r \hat {u}_a\right) = A_a,
\end{equation} 
and
\begin{align}
   \hat{u}_{m}=u_{c}(z,t), \quad \hat{v}_{m} = 0\;\;\;  \quad at \quad r &= 1, \label{uhat_bc_wall}\\
     \partial_r \hat{u}_a = 0, \quad \hat{v}_{a} = 0\;\;\;  \quad at \quad r &= 0,\label{uhat_bc_centreline}\\
        \partial_r \hat {u}_m = \Pi_\mu\partial_r \hat {u}_a, \quad \hat {u}_a = \hat {u}_m  \quad at   \quad r &= d,\label{uhat_bc_interface}
\end{align}
where $A_m$ and $A_a$ are determined by the flow rates $2\pi Q_i$:
\begin{align}\label{int_cons}
  \quad \int_{0}^{d} \hat{u}_a \,rdr  = Q_a , \quad \int_{d}^{1} \hat {u}_m \,rdr = Q_m.
\end{align}
Once $d$ and $Q_i$ are obtained by solving the WRIBL equations, given in the next sub-section, then $\hat{u}_i$ can be calculated to obtain the leading axial velocity profile at any position in the airway. Furthermore, the leading radial-velocity profile can be obtained by integrating the continuity equation \eqref{continuity}:
\begin{align}\label{v-comp-velocity}
  \quad \hat {v}_a = -\frac{1}{r}\int_{0}^{r} \partial_z\hat{u}_a rdr ,\quad \hat {v}_m = \frac{1}{r}\int_{r}^{1} \partial_z\hat{u}_m rdr
\end{align} 
In this manner, the leading-order, incompressible velocity field $(\hat{u}_i,\hat{v}_i)$ is obtained. It is used to (i) evolve particles in the air-phase, and (ii) visualize the flow by plotting streamlines obtained from contours of the streamfunction $\Psi_i$:
% \begin{align}\label{streamlines}
% \quad \Psi_m = \int_{r}^{1} \hat{u}_m \,r dr, \quad
% \Psi_a = \int_{0}^{r} \hat{u}_a \,r dr 
% \end{align}
\begin{align}\label{streamlines}
\quad \Psi_a = \int_{0}^{r} \hat{u}_a \,r dr, \quad
\Psi_m = \Psi_a|_d+\int_{d}^{r} \hat{u}_m \,r dr. 
\end{align}

Regarding the velocity corrections $u'_i$, equation \eqref{int_cons} implies that they must satisfy the following guage conditions, which will be used to eliminate $u'_i$ from the weighted-averaged equations.
\begin{align}\label{prime_int_conts}
  \quad \int_{0}^{d} u^{\prime}_a \,rdr  = 0 , \quad \int_{d}^{1}  u^{\prime}_m \,rdr = 0
\end{align}

\subsubsection{Weighted-residuals}
Denoting the boundary layer equations for the two phases in \eqref{BLE} by $BLE_i$, we obtain averaged equations by evaluating the residual $\langle BLE|w \rangle$ where the inner product is defined as $\langle p|q \rangle = \Pi_\mu \int_0^d{p_a q_a  r\,dr} + \int_d^1{p_m q_m r\,dr}$ and $w_i$ are weight functions. If naive radial averaging is performed with $w_i=1$, then the velocity correction $u'_i$ will arise in the residual via the leading-order, radial viscous diffusion term. One will then have to either calculate $u'_i$ or suffer prominent errors that can produce unphysical solutions \citep{Kalliadasis}. The ingenuity of the WRIBL approach \citep{Kalliadasis,Dietze2013} is to use a weight function which eliminates the correction from the residual (to $\mathcal{O}(\epsilon^2))$.

Multiple choices for the weights are possible. We follow \citet{Dietze2013,Dietze2015} and use two combinations. First, we choose $w_i$ so that the pressure-gradient term is cancelled out and an evolution equation for $Q_i$ is obtained. The defining equations for this weight function are
\begin{align}\label{weight_function}
  \quad \frac{1}{r}\partial_r \left(r\partial_r {w}_m\right) = 1, \quad \frac{1}{r}\partial_r \left(r\partial_r {w}_a\right) = C_a,
\end{align} 
  with the homogeneous version of the $\hat{u}_i$ boundary conditions:
  \begin{align}\label{weight_bc}
   {w}_{m}|_{r=1}=0, \quad \partial_r {w}_a|_{r=0} = 0, \quad \partial_r {w}_m|_{r=d} = \Pi_\mu\partial_r {w}_a|_{r=d}, \quad {w}_a|_{r=d} = {w}_m|_{r=d}.
\end{align}
$C_a$ in \eqref{weight_function} follows from applying
\begin{align}\label{wf_int_conts1}
  \int_{0}^{d} w_a rdr  = -\int_{d}^{1} w_m rdr 
\end{align}

The second weight function, denoted by $\Tilde{w}_i$, is chosen to yield a diagnostic equation for the pressure at the interface, $p_i|_d$. This equation will be used to enforce axial pressure boundary conditions. The defining equations for $\Tilde{w}_i$ are
\begin{align}\label{weight2_function}
  \quad \frac{1}{r}\partial_r \left(r\partial_r \Tilde{w}_m\right) = 1, \quad \frac{1}{r}\partial_r \left(r\partial_r \Tilde{w}_a\right) = \Tilde{C}_a,
\end{align}
with the same boundary conditions as $w_i$ but with
\begin{align}\label{wf_int_contsp}
  \int_{0}^{d} \Tilde{w}_a rdr  = \int_{d}^{1} \Tilde{w}_m rdr,
\end{align}
instead of \eqref{wf_int_conts1}. 

 On applying the velocity decomposition \eqref{vel_decomp}, along with \eqref{prime_int_conts}, the corrections $u'_i$ cancel out from the residual $\langle BLE|w \rangle$ at $\mathcal{O}(\epsilon^2)$, provided we neglect $\mathcal{O}(\Rey_i \epsilon^2)$ inertial terms, as is appropriate for low to moderate Reynolds numbers \citep{Dietze2015}. 

Note that our choice of weight functions differs from that in \citet{Dietze2015}. Our final WRIBL equations, listed below, have the same form as those in \citet{Dietze2015}, and the coefficients also match after rescaling to account for the difference in weight functions. We have also checked that the predictions of our WRIBL equations match those of \citet{Dietze2015}, as illustrated by figure~\ref{fig:validation} in Appendix~\ref{app:validate}. 

\subsubsection{WRIBL equations}
Two exact mass-conservation equations follow from integrating the continuity equation \eqref{continuity} in the radial direction, in each phase, and using the kinematic boundary condition \eqref{kinematic-gov}:
% \begin{equation}\label{mass_balance}
%     \partial_z Q_i - \delta_i  d \partial_t d = 0
% \end{equation}
\begin{align}
     d \partial_t d &= \partial_z Q_m, \label{mass_balance_d}\\
       \partial_z Q_m &+ \partial_z Q_a = 0 \label{mass_balance_q}
\end{align}
where, alternately, the latter could be replaced by $ d\partial_t d = -\partial_z Q_a$.

The evolution equation for the flow rate follows from evaluating the weighted-residual with respect to $w_i$:
\begin{multline}\label{wribl_q_eq}
 \frac{\mu_i}{\mu_m}Re_i\biggl(S_{ij}\partial_t Q_j+S_{ic}\partial_t u_c+F_{ijk} Q_j{\partial_z Q_k} +F_{ijc} Q_j\partial_z u_c+F_{icj} u_c\partial_z Q_j+ F_{icc}u_c\partial_zu_c\\+G_{ijk}Q_jQ_k\partial_z{d}+G_{icj}u_cQ_j\partial_z{d}+G_{icc}u_cu_c\partial_z{d}\biggr)\\
 =-Ca\left(\partial_z\kappa+\partial_z\kappa_p\right)Iw_a+\Pi_{\mu}C_aQ_a+Q_m+J_{j}Q_{j}(\partial_z {d})^2+J_{c}u_{c}(\partial_z {d})^2\\+K_{j}\partial_z Q_{j}\partial_z {d}+K_{c}\partial_z u_{c}\partial_z {d}
 +L_{j}Q_{j}\partial_z^2 {d}+L_{c}u_{c}\partial_z^2 {d}\\+M_j\partial_z^2Q_j+M_c\partial_z^2u_c-u_c\partial_r{w_m}|_{1}+\Pi_{\mu} G_{a}sin(\omega t)Iw_a,
\end{multline}
 where Einstein summation notation has been used on the phase index. The coefficients, $S_{ij},S_{ic},F_{ijk},...,M_c$ are functions of $d$ alone. The integrals of $\hat{u}$ and $w_i$, which yield these coefficients, are involved and the corresponding calculations are performed using the Python symbolic-computing library SymPy \citep{sympy}. 
 
 Note that, for convenience of notation, we have written the WRIBL equation \eqref{wribl_q_eq} after rescaling the axial coordinate with $R$; thus the length-scale ratio $\epsilon$ does not appear here. This new axial scaling will be used to present all subsequent equations and figures.

 Equations \eqref{mass_balance_d} to \eqref{wribl_q_eq} form a closed system for $d$, $Q_m$ and $Q_a$. Integrating \eqref{mass_balance_q} shows that the total flow rate, $Q_t(t) = Q_a+Q_m$, is spatially invariant. Therefore, given $Q_t$ as an input, one can replace $Q_a$ by $Q_t-Q_m$ and solve \eqref{mass_balance_d} and \eqref{wribl_q_eq} for $d$ and $Q_m$. Here, however, we do not impose $Q_t$. Rather, we impose the background pressure-gradient, so that the net pressure difference across the computational domain, of length $L$, must be be $G_{a}L\,sin(\omega t)$ in the air and zero in the mucus. The deviation from this applied pressure-gradient, $p_i$, must therefore integrate to zero over the domain:
 \begin{equation}\label{press_constraint}
 \int_{0}^{L/R } \partial_z p_i dz = 0
 \end{equation}

To apply this pressure constraint, we require the diagnostic equation for pressure, which is obtained by evaluating the weighted-residue with respect to $\Tilde{w}_i$:
\begin{multline}\label{wribl_p_eq}
 \frac{\mu_i}{\mu_m}Re_i\biggl(\Tilde{S}_{ij}\partial_t Q_j+\Tilde{S}_{ic}\partial_t u_c+\Tilde{F}_{ijk} Q_j\partial_z Q_k +\Tilde{F}_{ijc} Q_j\partial_z u_c+\Tilde{F}_{icj} u_c\partial_z Q_j+\Tilde{F}_{icc}u_c\partial_zu_c\\
 +\Tilde{G}_{ijk}Q_jQ_k\partial_z{d}+\Tilde{G}_{icj}u_cQ_j\partial_z{d}+\Tilde{G}_{icc}u_cu_c\partial_z{d}\biggr)\\
 =-2\Pi_{\mu}\partial_z p_a |_d\Tilde{I}w_a+Ca\left(\partial_z\kappa+\partial_z\kappa_p\right)\Tilde{I}w_a+\Pi_{\mu}\Tilde{C}_aQ_a+Q_m+\Tilde{J}_{j}Q_{j}(\partial_z {d})^2+\Tilde{J}_{c}u_{c}(\partial_z {d})^2\\+\Tilde{K}_{j}\partial_z Q_{j}\partial_z {d}+\Tilde{K}_{c}\partial_z u_{c}\partial_z {d}
 +\Tilde{L}_{j}Q_{j}\partial_z^2 {d}+\Tilde{L}_{c}u_{c}\partial_z^2 {d}\\+\Tilde{M}_j\partial_z^2Q_j+\Tilde{M}_c\partial_z^2u_c-u_c\partial_r{\Tilde{w}_m}|_{1}+\Pi_{\mu} G_{a}sin(\omega t)\Tilde{I}w_a.
\end{multline}

As is customary when using thin-film models to describe droplets, rivulets, and other situations involving regions devoid of liquid \citep{oron1997long,ghatak1999dynamics,RuyerQuil2023droplets}, we employ a precursor-film approach to capture the mucus-depleted zones of the airway. To wit, we introduce a disjoining pressure via the term $\kappa_p = \frac{h^6_0}{(1-d)^9}$ in \eqref{wribl_q_eq} and \eqref{wribl_p_eq}. With $h_0 = 5 \times 10^{-4}$, we obtain a precursor film thickness below 0.01 ($d>0.99$) in the depleted zones; the fluid drained out of these zones accumulates in annular humps whose shape, we have verified, is very close to that of equilibrium unduloids (of the same volume). Reducing the precursor film's thickness even further does not affect our results but significantly increases the computational cost.

The coefficients of \eqref{wribl_q_eq} and \eqref{wribl_p_eq} are stored in text files and imported when performing numerical simulations, as done in the Jupyter notebook associated with figure~\ref{fig:validation} in Appendix \ref{app:validate}.

\subsubsection{One-way coupled WRIBL equations}

Given that air has a much smaller viscosity and density than mucus, it is natural to check whether the simplified model obtained in the limit of $\Pi_{\rho}$, $\Pi_{\mu} \ll 1$ provides a good approximation of the two-phase flow. In this limit, the mucus film evolves independently of the air, with a stress-free surface. Such a passive-core approximation is routinely made when studying gas-liquid flows in which the gas flow is weak. While this approximation is certainly inappropriate for the upper airways, where the turbulent airflow is strong enough to rip off mucus droplets from the film, it could work well in the middle airways where the airflow is much weaker.

The WRIBL model for the mucus phase in the passive-core limit was analysed by \citet{Dietze2015}. Here, we additionally derive the WRIBL equation for the air phase, because we need the air velocity field in order to study aerosol transport. In the limit of infinite viscosity contrast, the air will experience the mucus film as a moving solid boundary. Thus, while the mucus film evolves independently, the airflow is modulated by the deformation of the film, \textit{i.e.}, we have a one-way coupled flow. Because the mucus velocity is typically much slower than that of the air, the major effect of the film is to alter the conduit for airflow.

The decoupled mucus flow is governed by the domain equations \eqref{continuity} and \eqref{BLE}, with $i = m$, the wall boundary-condition \eqref{no_pen_no_slip}, the kinematic condition \eqref{kinematic-gov}, and the free surface conditions obtained by setting $\Pi_\mu = 0$ in \eqref{tangential} and \eqref{normal}. (The smallness of $\Pi_{\rho}$ is used implicitly when we set $\Pi_\mu p_a$ to zero, in \eqref{normal}; $\Pi_{\rho}$ controls the relative strength of air pressure variations, in cases where inertia is significant.) We then apply the WRIBL method to the mucus phase by performing the weighted integral from $r=d$ to $r = 1$. The leading-order mucus velocity is determined by the equations for $\hat{u}_m$ in \eqref{u_hat} to \eqref{int_cons} but now with $\Pi_\mu = 0$. Similarly, the weight function is defined by the equations for $w_m$ in \eqref{weight_function} and \eqref{weight_bc}, with $\Pi_\mu = 0$. The resulting WRIBL model for the mucus-film is as follows:
% If $\Pi_{\mu}$ and $\Pi_{\rho}$ $\to {0}$ then the interface coupling boundary condition Eq.(\ref{normal},\ref{tangential}) reduced to 
% \begin{equation}\label{normal}
%  p_m  =  -Ca (\partial_z \kappa) + 2\epsilon^2\left(\partial_r v_m- \partial_r u_m \partial_z d \right) 
%  \end{equation}
 
%  \begin{equation}\label{tangential}
%    \partial_r u_m = 2\epsilon^2 \partial_z d\left(\partial_z u_m- \partial_r v_m\right)-\epsilon^2 \partial_z v_m 
%  \end{equation}
%  \begin{align}
%    \hat{u}_{m}=u_{c}(z,t), \quad \hat{v}_{m} = 0\;\;\;  \quad at \quad r &= 1,\\
%    \partial_r \hat {u}_m = 0,   \quad at   \quad r &= d,
% \end{align}
% which ultimately leads to a simplified version of WRIBL equation for the mucus phase considering air as passive.
\begin{equation}\label{kbc_mucus}
   d \partial_t d = \partial_z Q_m  
\end{equation}
\begin{multline}\label{wribl_q_eq_pas}
Re_m\biggl(\overline{S}_{mm}{\partial_t Q_m}+\overline{S}_{mc}{\partial_t u_c}+\overline{F}_{mmm} Q_m\partial_z Q_m +\overline{F}_{mmc}Q_m\partial_z u_c+\overline{F}_{mcc}u_c\partial_z u_c+\\
\overline{F}_{mcm} u_c\partial_z Q_m+\overline{G}_{mmm}Q_mQ_m\partial_z{d}+\overline{G}_{mcm}Q_m u_c\partial_z{d}+\overline{G}_{mcc}u_c u_c\partial_z{d}\biggr)\\
 =Ca\left(\partial_z\kappa+\partial_z\kappa_p\right)\overline{Iw}_m+Q_m+\overline{J}_{m}Q_{m}(\partial_z {d})^2+\overline{J}_{c}u_{c}(\partial_z {d})^2+\overline{K}_{m}\partial_z Q_{m}\partial_z {d}\\+\overline{K}_{c}{\partial_z u_{c}}{\partial_z {d}}
 +\overline{L}_{m}Q_{m}{\partial_z^2 {d}}+\overline{L}_{w}u_{c}{\partial_z^2 {d}}+\overline{M}_m{\partial_z^2Q_m}+\overline{M}_c{\partial_z^2u_c}-u_c\partial_r{\overline{w_m}}|_{1}
\end{multline}

Equations \eqref{kbc_mucus} and \eqref{wribl_q_eq_pas} must be solved simultaneously to determine $Q_m$ and $d$, which will then be used as inputs to calculate $Q_a$ from the WRIBL equation in the air phase. The leading-order air velocity $\hat{u}_a$ is determined by \eqref{u_hat} and \eqref{int_cons} along with the boundary conditions $\partial_r \hat{u}_a|_0 = 0$ and $\hat{u}_a|_d = \hat{u}_m|_d$. The weight function $w_a$ satisfies the homogeneous version of these boundary conditions, in addition to \eqref{weight_function}. Performing the weighted integral over the air phase, from $r=0$ to $r = d$, yields the WRIBL equation for the airflow:
% that predicts the response of the airflow to the independent-evolution of the mucus film:

% As next step we go to solve the core air-phase. Here we solve the air phase completely independently. the boundary conditions has been used are as follows. 
% \begin{align}
%      \partial_r \hat{u}_a = 0, \quad \hat{v}_{a} = 0\;\;\;  \quad at \quad r &= 0,\\
%     \hat {u}_a = {u}_d  \quad at   \quad r &= d,
% \end{align}
% Here $u_{d} = \hat u_m|_{r = d}$.
\begin{multline}\label{wribl_q_eq_air}
Re_a\biggl(\overline{S}_{aa}{\partial_t Q_a}+\overline{S}_{ad}{\partial_t u_d}+\overline{F}_{aaa} Q_a{\partial_z Q_a} +\overline{F}_{aad}Q_a{\partial_z u_d}+\\
\overline{F}_{ada} u_d{\partial_z Q_a}+\overline{F}_{add}u_d\partial_z u_d+\overline{G}_{aaa}Q_aQ_a{\partial_z{d}}+\overline{G}_{ada}u_dQ_a{\partial_z{d}}+\overline{G}_{add}u_du_d{\partial_z{d}}\biggr)\\
 =-\partial_z p_a|_d\overline{Iw}_a+\overline{C}_aQ_a+\overline{J}_{a}Q_{a}(\partial_z {d})^2+\overline{J}_{d}u_{d}(\partial_z {d})^2+\overline{K}_{a}{\partial_z Q_{a}}{\partial_z {d}}+\overline{K}_{d}{\partial_z u_{d}}{\partial_z {d}}\\
 +\overline{L}_{a}Q_{a}{\partial_z^2 {d}}+\overline{L}_{d}u_{d}{\partial_z^2 {d}}+\overline{M}_a{\partial_z^2Q_a}+\overline{M}_d{\partial_z^2u_d}+G_{a}sin(\omega t)\overline{Iw}_a-u_d\partial_r{\overline{w_a}}|_{d}.
\end{multline}
This equation, which has the air flow rate $Q_a$ and the pressure at the interface $p_a|_d$ as unknowns, must be solved along with the overall mass balance equation \eqref{mass_balance_q}.

\subsubsection{Numerical solution}

To solve the fully coupled WRIBL equations, we first use \eqref{mass_balance_q} to substitute $Q_m = Q_t(t)-Q_a$ in \eqref{wribl_q_eq} and \eqref{wribl_p_eq}. The latter pressure-equation is then integrated over the domain and the pressure-constraint \eqref{press_constraint} is applied to obtain an ordinary differential equation (ODE) for $Q_t(t)$ of the form
\begin{equation}\label{Qt_ode}
d_t Q_t = \Phi (d,Q_a,Q_t,u_c)
\end{equation}
Equations \eqref{mass_balance_d}, \eqref{wribl_q_eq} and \eqref{Qt_ode} thus become a closed system for $d$, $Q_a$ and $Q_t$. 

Numerical simulations are performed by discretizing space using a second-order central-difference scheme. The resulting system of ODEs is integrated in time using the stiff, adaptive time-stepping, LSODA solver \citep{hindmarsh1995algorithms} provided by the function solve{\textunderscore}ivp, of the Python library SciPy \citep{scipy}.      

Turning to the one-way coupled WRIBL model, the equations for the mucus film, namely \eqref{kbc_mucus} and \eqref{wribl_q_eq_pas}, are solved first to obtain $d$ and $Q_m$. Here too we employ second-order central-differencing in space, followed by the LSODA method for stepping in time. To obtain $Q_a$, we need only solve for $Q_t$, since $Q_a = Q_t(t)-Q_m$; on substituting this relation into the one-way coupled WRIBL equation for air, \eqref{wribl_q_eq_air}, integrating over the domain, and applying the pressure constraint \eqref{press_constraint}, we obtain an ODE for $Q_t$. So, rather than simultaneously solving two partial differential equations (PDEs) and one ODE, as required by the fully coupled WRIBL model, we solve just two PDEs first and then separately solve a single ODE. 
% Therefore, as expected, simulations of the one-way coupled WRIBL model are faster. 

All our simulations are performed with periodic boundary conditions in the axial $z$ direction. The length of the computational domain $L$ is chosen to match the wavelength of the fastest growing mode of the Rayleigh-Plateau instability, $\Lambda_{RP} = 2\pi 2^{1/2} d_0 R$. (This inviscid prediction of \citet{Rayleigh1892wavelength} works very well even for viscous films \citep{Dietze2015}.) An evenly-spaced spatial grid of 500 points is found to be sufficient for obtaining grid-independent solutions.
% We also perform simulations on longer domains with $L = 4 \Lambda_{RP}$ to test whether calculations based on just the fastest growing mode are able to capture the dependence of the extent of mucus-depleted zones on the initial film thickness $1-d_0$. 

% When solving for the mucus film alone, a spatial resolution of 300 grid point suffices for the long domain of $L = 4 \Lambda_{RP}$. In contrast, when solving for both the air and mucus, we need up to 500 grid points just for the domain of $L = \Lambda_{RP}$.

\subsection{Particle motion}

To track the motion of airborne particles, we adopt the point-particle approximation and use the Maxey-Riley equation, simplified for tiny dense particles \citep{ravichandran-rev2017}. In addition, we include Brownian forces, in order to describe a wide range of particles, from Brownian to inertial. The particle's position $\boldsymbol{X}_p$ and velocity $\boldsymbol{V}_p$ are thus governed by the following stochastic differential equations (SDEs), which are often used for studying particle transport in the lungs \citep{tsuda2013particle}: 
\begin{align}
  d\boldsymbol{X}_p &= \boldsymbol{V}_p\, dt \label{eqn_cm_part}\\
    St\,d\boldsymbol{V}_p &= \left[\boldsymbol{u}(\boldsymbol{X}_p,t)-\boldsymbol{V}_p\right]dt+\sqrt{\frac{2}{\Pen}}\, d\boldsymbol{W},
    \label{eqn_langevin}
\end{align}
where $\boldsymbol{u}$ is the air-velocity vector which must be obtained at the instantaneous position of the particle, $\boldsymbol{X}_p$. 
Equation \eqref{eqn_langevin} equates the change in the particle's momentum to the sum of the Stokes drag force (proportional to the slip velocity, $\boldsymbol{u}-\boldsymbol{V}_p$) and the Brownian force. The latter is modelled by $d\boldsymbol{W}$, a vector of independent increments of the Wiener process \citep{gardiner}. The impact of Brownian forces on the particle's motion varies inversely with the P\'eclet number ($\Pen$), which is the ratio of the timescales of Brownian diffusion to convection by the airflow. The importance of inertial effects increases with the Stokes number ($St$), which is the ratio of the timescale of the particle's inertial relaxation to the time-scale over which the flow changes (as viewed by the convected particle). In our study, $\Pen$ and $St$ are not independent but vary together with the particle diameter $d_p$, as follows:
% The strengths of Brownian and inertial effects on the particle's motion is determined by its P\'eclet ($\Pen$) and Stokes ($St$) numbers: 
\begin{equation} \label{Pe-St}
    \Pen = \frac{R U^\star}{D} \; \mathrm{with} \; D = \frac{k_B T}{3\pi\mu_a d_p}, \; \mathrm{and} \; St = \frac{d_p^2 \rho_p U^\star}{18 \mu_a R},
\end{equation}
where $D$ is the Einstein diffusivity of the particle, $k_B$ is the Boltzmann constant, $T = 298 K$ is the temperature, and $\rho_p$ is the particle's density. 
% \textcolor{red}{$U_\mathrm{max} = \mathrm{max}_t\left[\langle u_a|_{d_\mathrm{min}}\rangle_r\right] U$} 
The air velocity scale $U^\star$ is chosen to be $U_\mathrm{max}/2$, where $U_\mathrm{max} = \mathrm{max}_t\left[2({ Q_a|_{d_\mathrm{min}}})/{d_{min}^2}\right] U$ is the maximum, over time, of the cross-section averaged air velocity below the mucus hump (the $z$ location where $d = d_\mathrm{min}$); it yields an estimate of the smallest convective time-scale during respiration. We fix the particle's density to that of water ($\rho_p = 1000$ \si{kg.m^{-3}}) and vary the diameter from $10^{-3}$ to 50 \si{\mu m}. Both $\Pen$ and $St$ increase strongly with $d_p$ (see table~\ref{table:particles}) so that the motion of the smallest particles is nearly Brownian while that of the largest is influenced by inertial effects, such as migration across curved streamlines. Intermediate-sized particles are affected the least by these forces and tend to follow the flow like tracers.

In solving \eqref{eqn_cm_part}-\eqref{eqn_langevin}, we use a two-step exponential integration scheme, which enables accurate simulations of small $St$ particles \citep{ireland}. The Wiener increment $d\boldsymbol{W}$ is updated only once every time step (it stays the same for both half-steps); thus, the Brownian term is integrated using the Euler-Marujama scheme \citep{ottinger}. This strategy, of using higher-order time-stepping for the drift terms of an SDE, is helpful when the relative importance of Brownian forces varies widely; it has been used, for example, to study Brownian elastic chains in turbulent flows (\citeauthor{Bagheri} \citeyear{Bagheri}; \citeauthor*{Singh2022} \citeyear{Bagheri}). 

\begin{table}
 \begin{center}
\phantom{~}\noindent
\begingroup
\renewcommand{\arraystretch}{1.2}
  \begin{tabular}{p{0.35\textwidth}>{\centering}p{0.1\textwidth}>{\centering}p{0.2\textwidth}>{\centering\arraybackslash}p{0.2\textwidth}}
         Parameter & Symbol & Minimum & Maximum\\[6pt]  
    diameter (\si{\mu m}) & $d_p$  & $0.001 $ & $50 $\\
    diffusion time (s) & $-$ &  $6.39$ & $3.2\times 10^5$\\
    P\'eclet number& $\Pen$ & $1\times 10^2 $ & $9\times 10^6$\\
    inertial relaxation time (s)& $-$&  $3\times 10^{-12}$  & $7\times 10^{-3}$\\
    Stokes number& $St$ & $9\times 10^{-11}$ & $2\times 10^{-1}$\\
\end{tabular}
\endgroup
\caption{Range of particle sizes considered in this study and the corresponding timescales and non-dimensional numbers. 
% The density of all particles is taken to be 1000 \si{kg.m^{-3}}.
}
\label{table:particles}
 \end{center}
\end{table}

The solution of the WRIBL model yields the air velocity field in axisymmetric cylindrical coordinates. We transform this field to three-dimensional Cartesian coordinates and then evolve particles in three-dimensions. The air velocity at the particle location is obtained from its value on the computational grid via linear interpolation. We consider the dilute limit and neglect inter-particle collisions. Thus, the motion of each particle is independent of all others in the airway; we use this fact to parallelize the numerical integration of the particles to accelerate simulations. We track up to 2000 particles in a given flow, for up to 30 breathing cycles. 

Adopting the simplest approach to studying deposition, we assume that a particle deposits when its surface makes contact with the film \citep{guha2008transport}. 
We therefore monitor the radial distance (along the $r$ coordinate of the airway) between the centre of mass of the particle and the mucus-air interface; if this distance falls below $d_p/2$, then we assume that the particle has collided with the mucus film and become instantly entrapped; the trapped particle is no longer evolved. Strictly speaking, we should monitor the shortest distance between the particle and the interface, but this would be computationally expensive. So, noting that axial variations of the film occur over distances much larger than the particle's radius, we replace the shortest distance by the radial distance. In keeping with our use of a precursor film to represent mucus-depleted sections of the wall, we treat all particles that collide with the precursor film (of thickness less than $0.01R$) as having deposited on the wall.

This model of particle transport, while simple, captures the key physical aspects of deposition in a mucus-lined airway: (i) size-dependent Brownian and inertial forces, (ii) possibility of deposition on mucus-depleted zones of the wall, and (iii) influence of spatially non-uniform airflow, due to the presence of protruding mucus humps. There are of course various directions in which the particle model can be improved. In particular, the largest particles considered here are 1/8 times the diameter of the airway and are thus not well-suited to be treated as point particles. Nonetheless, we include these particles in our study, with the aim of providing a first approximation to their deposition behaviour. Indeed, the influence of finite-size effects (such as shear-gradient-induced lift and wall-induced lift) on the particle's motion scale with the fluid to particle density ratio \citep{guha2008transport,lift2016}; here $\rho_p/\rho_a = 10^{-3}$ and so the effects of finite-size will be weak in comparison to the effects of particle inertia (such as centrifugation). The latter will be shown to play an important role in the presence of a non-uniform mucus film.
% , and we expect these conclusions regarding inertial particles to remain valid even for a more complex model of their motion.

\section{How cilia and air affect the mucus film}\label{sec:cilia-air}

We shall first examine how the mucus film---its distribution and flow---is affected by cilia and air. Then, in \S~\ref{sec:depleted}, we will quantify the extent of mucus depleted zones. The understanding of the flow gained in these sections will aid our study of particle deposition, which will be taken up in \S~\ref{sec:particles}.

All our simulations use the parameters given in table~\ref{table:simulation-parameter}, which are typical for mucus and air in a middle airway (we focus on generations 14-16). Our results are not sensitive to the precise values of the parameters, as demonstrated in the \href{https://bighome.iitb.ac.in/index.php/s/6p6PHMqSKYHD6sx}{ supplementary material} which reproduces select figures for a different value of the viscosity ratio and interfacial tension. 

Throughout this work, we set the length of the computational domain to match the wavelength of the fastest growing mode of the Rayleigh-Plateau instability, i.e., $L = \Lambda_{RP}= 2\pi 2^{1/2} d_0 R$ \citep{Rayleigh1892wavelength,Dietze2015}. Furthermore, we only consider films with an initial thickness $1-d_0$ sufficiently small to remain open without forming a liquid bridge, \textit{i.e.}, with $(1-d_0^2) \Lambda_{RP}/R < 1.73$ \citep{Dietze2015,everett1972model}. In the present section, representative results for the air-mucus flow are shown for the value of $d_0 = 0.885$. The simulations are initialized with a slightly perturbed interface, $d = d_0 + 10^{-3}\,\mathrm{cos}(2 \pi z R/\Lambda_{RP})$. Random perturbations will be applied later, in \S~\ref{sec:depleted}, while analysing the extent of mucus-depleted zones.

\subsection{Cilia translates the mucus film}\label{sec:cilia}

\begin{figure}
\includegraphics[width=1.0\textwidth]{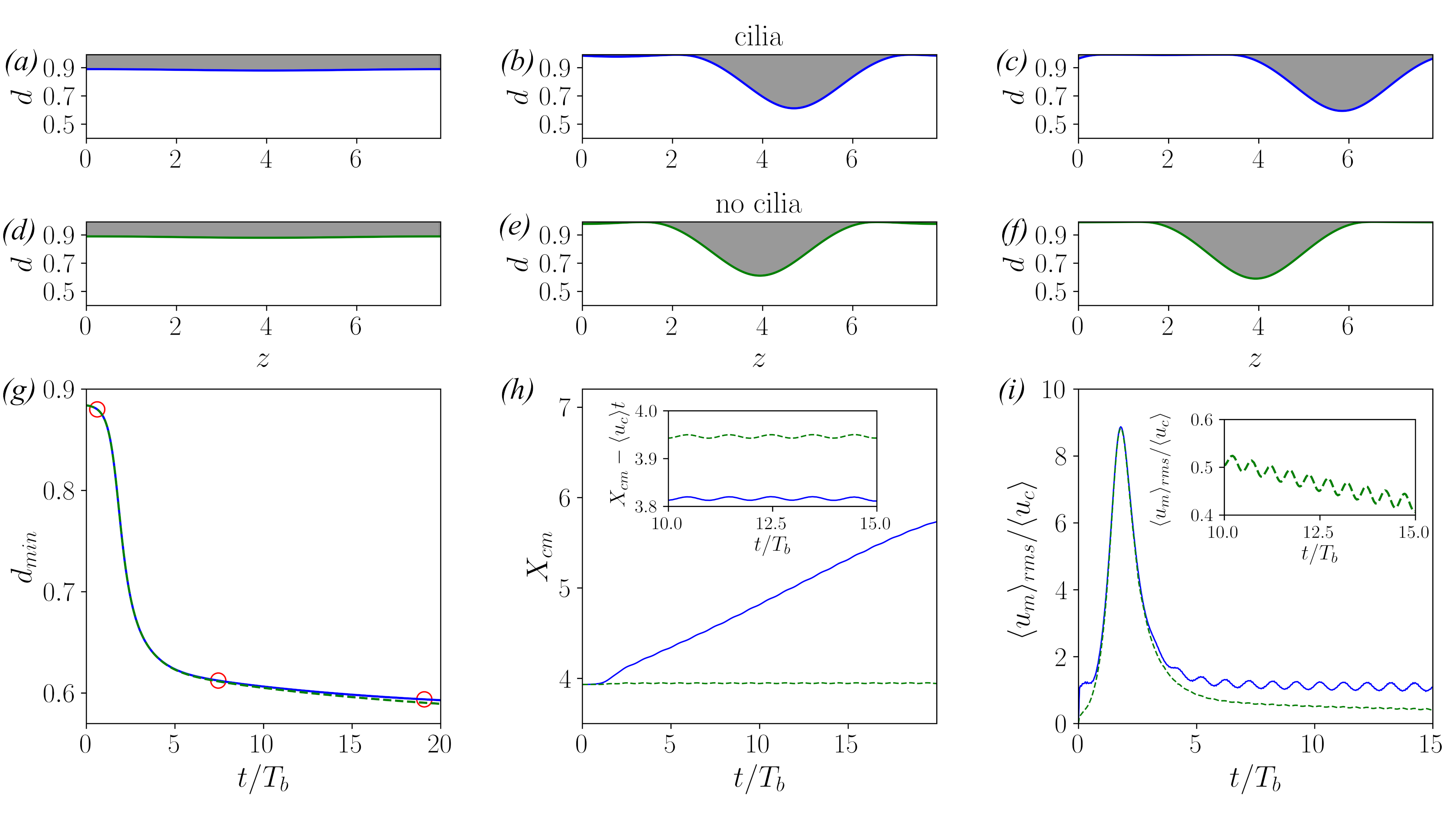} 
\caption{\label{fig:cilia-interface} 
Cilia translates the mucus film without altering the capillary-driven dynamics and the consequent emergence of humps and depleted zones. The top two rows present snapshots of the evolution of the film, with cilia (\textit{a}-\textit{c}) and without cilia, \textit{i.e.}, with $u_c = 0$ (\textit{d}-\textit{f}). Panel (\textit{g}) compares the growth of the hump by tracing the evolution of the minimum of $d(z,t)$. Panel (\textit{h}) compares the time-traces of the centre of mass of the entire film, clearly showing that cilia causes the film to translate; the inset subtracts out the net translation $\langle u_c \rangle t$  and reveals tiny lateral air-induced oscillations.
% , which, being caused by respiratory airflow, are present even in the case without cilia. 
Panel (\textit{i}) presents the time-trace of the spatially-averaged root-mean-square mucus velocity, normalized with the mean cilia velocity; the inset zooms into the curve for the no-cilia case.}
% ; the inset zooms into the same curves to reveal the structure of the air-induced oscillations.  }
\end{figure}

The effect of cilia on the film profile is illustrated in figure~\ref{fig:cilia-interface}. Snapshots of the evolution of the interface with ciliary forcing (via the metachronal velocity $u_c$ imposed at the wall, as per \eqref{cilia_bc}) are compared with those without ciliary forcing (setting $u_c = 0$), in panels (\textit{a}-\textit{c}) and (\textit{d}-\textit{f}). The same unduloid-shaped hump is formed in both cases, but the hump translates in the presence of cilia. The growth of the hump, visualized by the time trace of the minimum position of the interface in figure~\ref{fig:cilia-interface}(\textit{g}), is indeed nearly identical in the two cases. In this and all subsequent plots, we present the temporal evolution in multiples of the non-dimensional breathing time-period $T_b$ (where the dimensional time period, $T_b R/U$ is 1 s). 

The cilia-induced translation shows up clearly in figure~\ref{fig:cilia-interface}(\textit{h}) which traces the location of the centre-of-mass of the film, $X_{cm}$; the translation is near-linear with a constant speed equal to the mean cilia speed $\langle u_c \rangle$. When this linear translation is subtracted out, as done in the inset, we find that tiny oscillations are superimposed on the linear translation; the oscillations are caused by respiratory airflow as evidenced by their presence in the case without cilia (for which we use $\langle u_c \rangle = 0$ in the inset).

These air-induced oscillations are also seen in the root-mean-square averaged velocity of mucus, 
% \textcolor{red}{$\langle u_m\rangle_\mathrm{rms} = [\int_0^L{ u_m^2 dz}]^{1/2}$}, 
 {$\langle u_m\rangle_\mathrm{rms} = \left[{L^{-1}}\int_0^L{\left({2 Q_m}/({1-d^2})\right)^2 dz}\right]^{1/2}$},
which is plotted in figure~\ref{fig:cilia-interface}(\textit{i}). 
% The inset zooms into a few oscillations and shows that, in the presence of cilia, every alternate oscillation has a smaller amplitude; this occurs during the half-period of breathing when the air flow is in the opposite direction to $\langle u_c \rangle$ and thus counteracts the ciliary transport of mucus. 
Here, the prominent peak in $\langle u_m\rangle_\mathrm{rms}$ corresponds to the relatively rapid capillary-driven flow of mucus that produces the hump. After the hump has formed (and the capillary forces due to axial and radial curvatures have balanced out) the interface profile remains unchanged while the cilia slowly transports the film. 

% Once the hump forms, the capillary forces due to the axial curvature balance those due to the radial curvature and the net capillary pressure-gradients become very weak. The long-term flow is therefore driven by cilia, which produces a velocity that is about five times smaller than that which occurs during hump formation. This observation helps explain why cilia has no effect of the formation of the hump: apart from acting in the horizontal rather than the radial direction, the cilia-induced velocity is much too weak to alter the capillary-driven flow by which mucus gathers into humps. 

\begin{figure}
\includegraphics[width=1.0\textwidth]{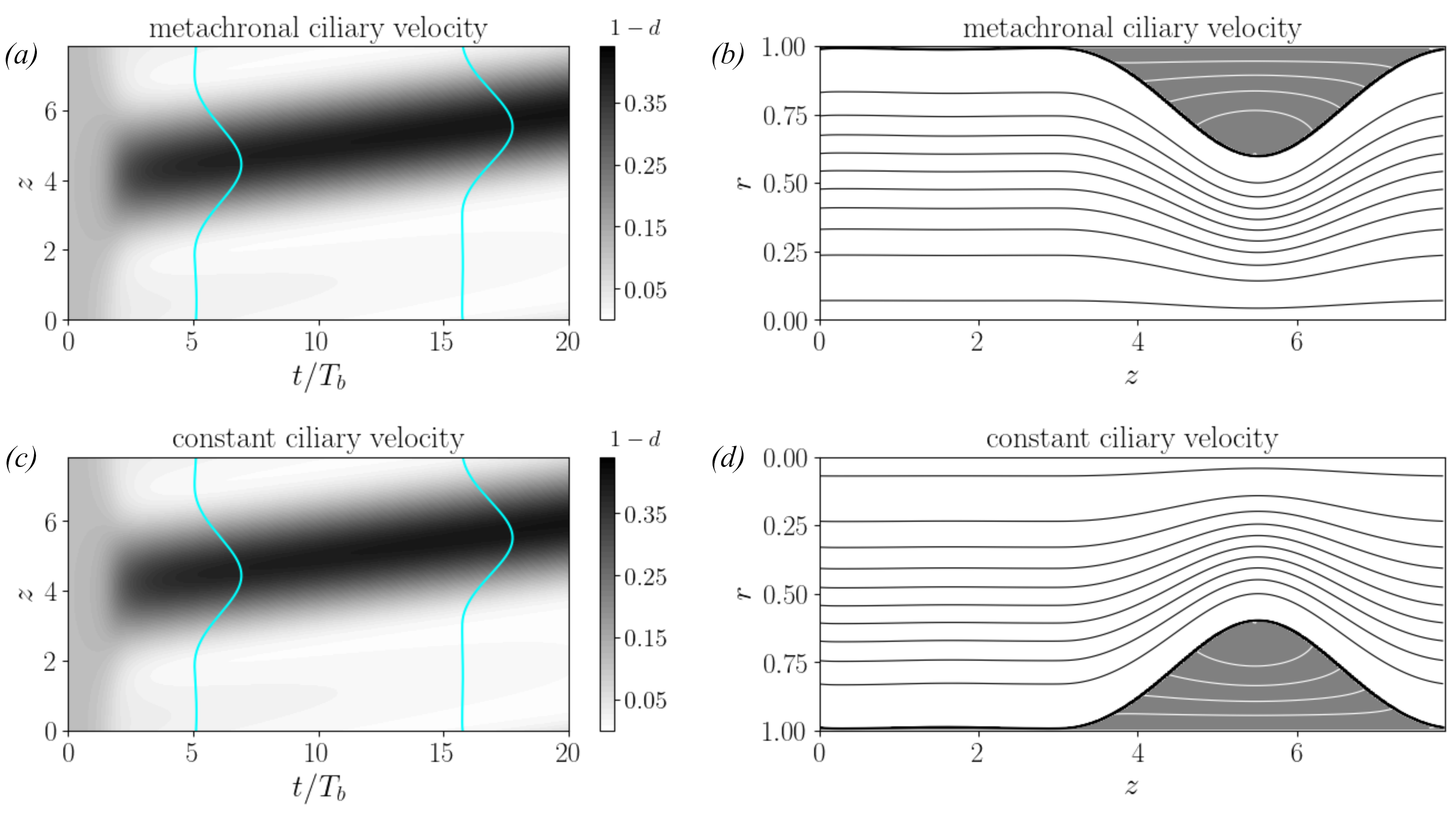}% Here is how to import EPS art
\caption{\label{fig:cilia_simple_kymo} The metrachronal cilia velocity \eqref{cilia_bc} has the same effect on the flow as the constant cilia velocity \eqref{cilia_bc_const}, as evidenced by the kymographs of the evolution of the film thickness $1-d(z,t)$ (panels \textit{a}, \textit{c}) and the snapshots of the streamlines in the air and mucus (panels \textit{b}, \textit{d}). 
% The top (bottom) row corresponds to the metachronal (constant) cilia velocity. 
The snapshot of the streamlines corresponds to the time of the second line profile drawn in the kymographs ($t \approx 15.6 \,T_b$).}
\end{figure}
\begin{figure}
\includegraphics[width=1.0\textwidth]{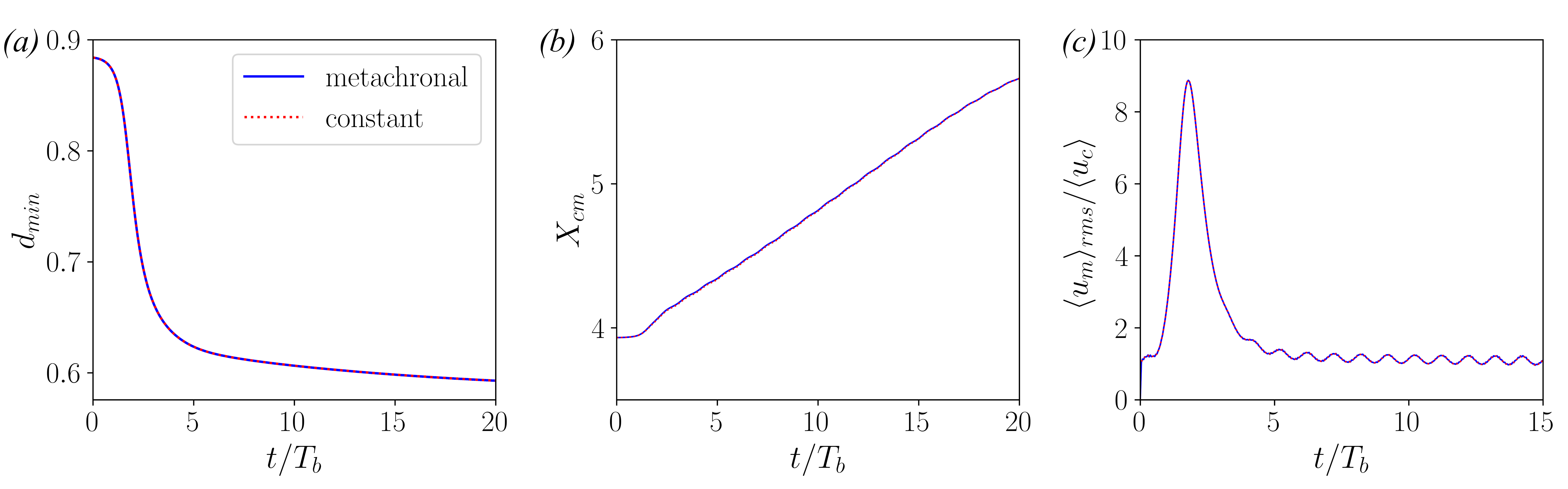}
\caption{\label{fig:cilia_simple} Comparison of the time traces of (\textit{a}) the minimum position of the interface, (\textit{b}) the centre-of-mass of the film, and (\textit{c}) the root-mean-square mucus velocity, for the metrachronal and constant cilia velocities (given by \eqref{cilia_bc} and \eqref{cilia_bc_const} respectively). The curves overlap in each panel, showing perfect agreement.}.     
\end{figure}

Since hump-formation is a one-time event, we will focus on the long term motion of the film for our study of particle deposition. The space-time kymograph of the interface profile in figure~\ref{fig:cilia_simple_kymo}(\textit{a}) emphasizes that this long-term motion is dominated by simple translation. In fact, if we replace the asymmetric metachronal wave form of $u_c$ in \eqref{cilia_bc} by just its mean, as given by the constant velocity Couette condition in \eqref{cilia_bc_const}, we obtain a near-identical translatory motion of the interface, as shown in figure~\ref{fig:cilia_simple_kymo}(\textit{c}). The streamlines in the air and mucus phases are also found to match very well (compare figures \ref{fig:cilia_simple_kymo}\textit{b} and \ref{fig:cilia_simple_kymo}\textit{d}). Further evidence for this equivalence is provided in figure~\ref{fig:cilia_simple} in which the time-traces of $d_{min}$, $X_{cm}$, and $\langle u_m\rangle_\mathrm{rms}$ for the constant cilia velocity are seen to match those for the metrachronal cilia velocity. Thus, the metachronal waveform of $u_c$ plays no role in the evolution of the interface and the net mucociliary transport---only the mean $\langle u_c \rangle$ matters (we have checked that this result is independent of spatial resolution by increasing the grid points from 300 to 1024). Hence, in all subsequent results, we shall use the constant cilia velocity boundary condition \eqref{cilia_bc_const}, both for simplicity and ease of computation (without the relatively rapid oscillations in $u_c$ we can take much larger time steps while integrating the WRIBL equations).

Of course, our conclusion regarding the unimportance of the metachronal waveform is valid only for the slow evolution of the film over long times scales; indeed, this result is unsurprising given that the cilia-beating time period is about three orders of magnitude smaller than the time scale with which mucus is transported across the airway (table~\ref{table:properties}). Another caveat is that we have considered a Newtonian mucus film, in which the dominant mechanism of momentum transport from the ciliary boundary is viscous diffusion, which effectively filters out the rapid oscillations of $u_c$. In contrast, for strongly viscoelastic mucus, present under diseased conditions, \citet{Dietze2023mucociliary} have shown that the metachronal waveform of $u_c$ produces a substantial reduction of the net mucociliary transport (in a flat, non-deforming mucus film). However, for healthy mucus, which is not as viscoelastic, this effect is marginal; indeed, the flow rate of a flat Newtonian film is shown to depend only on $\langle u_c \rangle$ \citep{Dietze2023mucociliary}. 

The cilia-induced translation of mucus is much slower than the respiratory airflow, as is evident in figure~\ref{fig:cilia-interface}(\textit{h}): the interface hardly moves in a single breath, \textit{i.e.}, in a time interval of $T_b$ (see also the corresponding time and velocity scales in table~\ref{table:properties}). So, the interface will appear to be stationary to air-borne particles which typically travel back and forth across the domain with each breath. Indeed, we see in figure~\ref{fig:cilia_simple_kymo}(\textit{b}) that the air streamlines bend around the mucus hump as if it were stationary. So, while ciliary transport of mucus is essential for the ultimate evacuation of trapped particles from the lungs, it cannot play a role in the deposition of particles on the mucus; mucociliary transport is simply too slow relative to the motion of airborne particles to produce any collisions that would not occur without cilia.

\subsection{Airflow does not alter the distribution of the mucus film}\label{sec:air}

\begin{figure}
\begin{center}
\includegraphics[width=.9\textwidth]{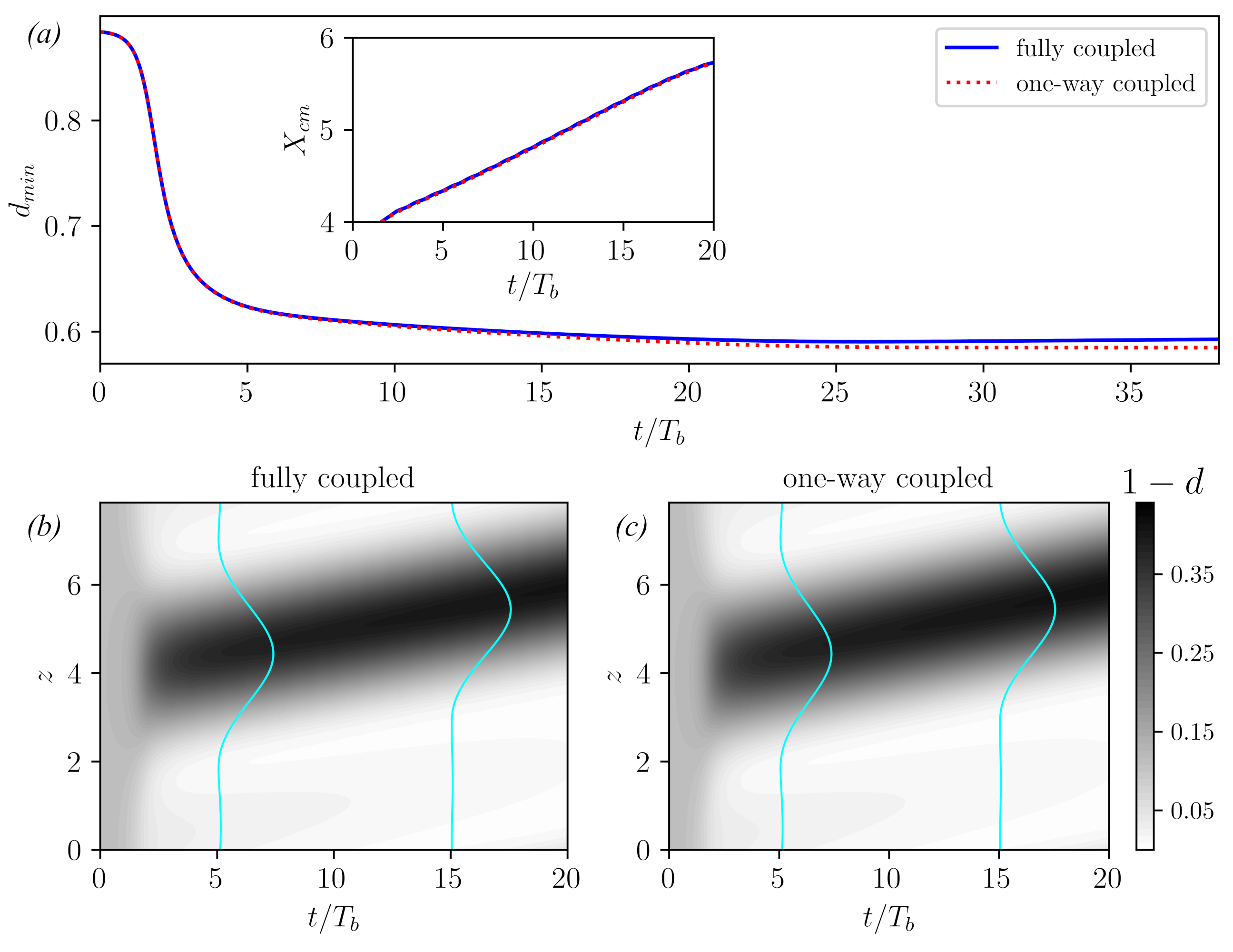}% Here is how to import EPS art
\caption{\label{fig:oneway-interface} 
Comparison of the interface evolution predicted by the fully-coupled and the one-way coupled WRIBL models; in the latter, air does not affect the flow mucus film. (\textit{a}) Evolution of the minimum interface position (main panel) and the centre of mass (inset). Kymographs of the film thickness are compared in panels (\textit{b-c}).
% ($t \approx 15 \,T_b$).
}
\end{center}
\end{figure}
\begin{figure}
\includegraphics[width=1.0\textwidth]{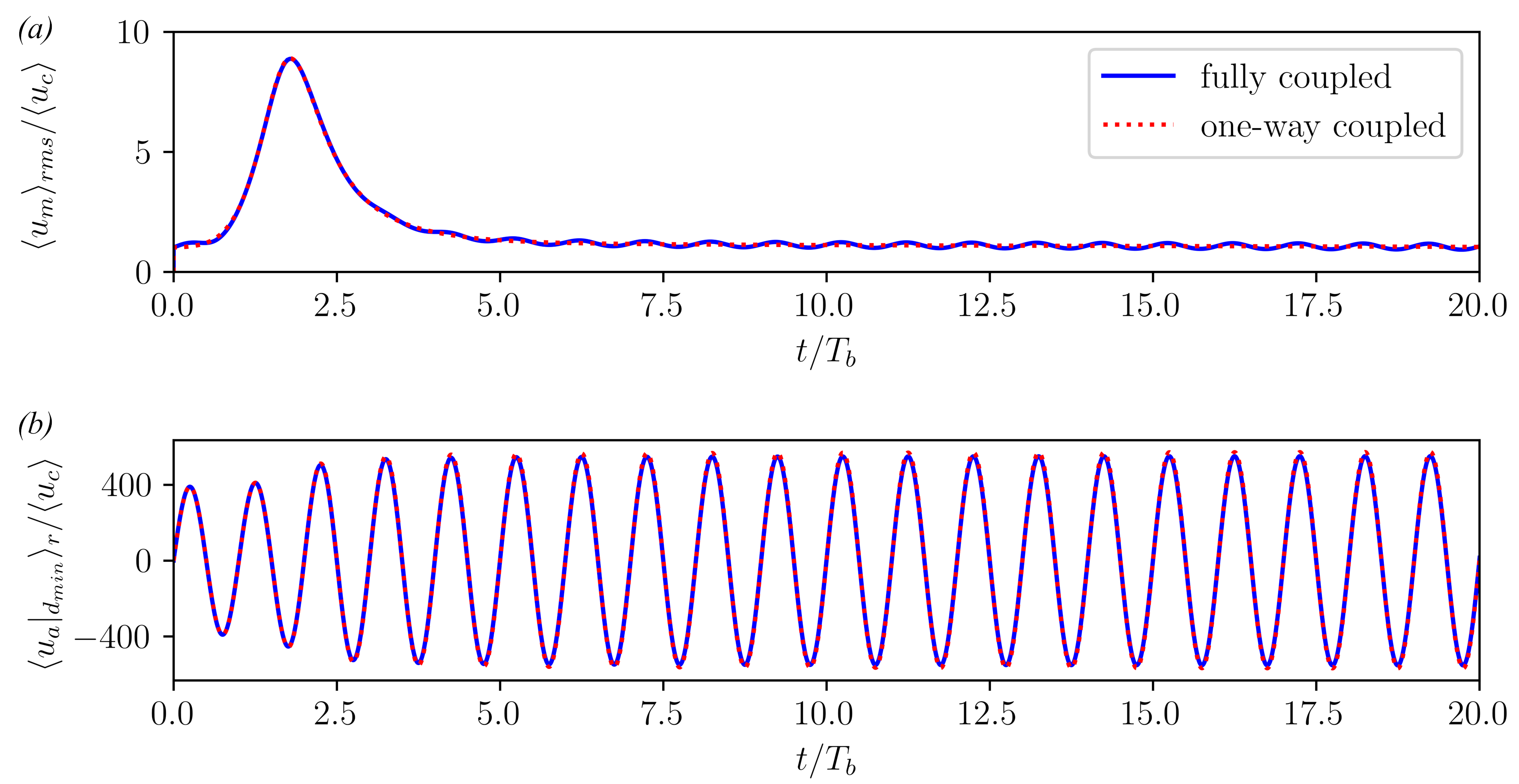}% Here is how to import EPS art
\caption{\label{fig:oneway-vel} Comparison of the evolution of averaged velocities in (\textit{a}) mucus and (\textit{b}) air, as predicted by the fully-coupled and the one-way coupled WRIBL models. Panel (\textit{a}) presents the spatially-averaged root-mean-square mucus velocity; note the absence of the small air-induced oscillations in prediction of the one-way coupled model, which ignores the influence of air on the mucus film. Panel (\textit{b}) presents the radial average of the air velocity just below the hump, i.e., below the thickest part of the interface given by $d_{min}$.}
\end{figure}

Next, we consider the effect of respiratory airflow on the film. The small oscillations in figure \ref{fig:cilia-interface}(\textit{i}) show that the flow of mucus is marginally modulated by the airflow. 
% The streamlines in the mucus, in figure \ref{fig:cilia_simple_kymo}(\textit{b}), also indicate that the top surface of the mucus film is dragged along with the air. 
This effect would obviously not be captured by the one-way coupled WRIBL model, in which the flow of air does not affect that of mucus. However, aerosol transport and deposition depend crucially on the mucus film's distribution (shape and size of humps and depleted zones) and not on its internal flow. Moreover, because the flow of mucus is much slower than that of air, the film will influence the airflow primarily via changes in its profile. The interface profile $d(z,t)$, and the flow of air around it, might well be captured by the one-way model even though the flow inside the mucus film is not exactly reproduced.

Figure~\ref{fig:oneway-interface} shows that the one-way coupled model does indeed capture the evolution of the interface almost perfectly. The time traces of $X_{cm}$ and the minimum of $d$ practically overlap (figure~\ref{fig:oneway-interface}\textit{a}), and the kymographs exhibit the same translating hump (figures~\ref{fig:oneway-interface}\textit{b}-\ref{fig:oneway-interface}\textit{c}). 
% The only discrepancy is the miniscule back and forth motion of the film caused by airflow, which is only revealed on subtracting out the cilia-induced translation, as done in the inset of \ref{fig:cilia-interface}(\textit{h}).
% by the time trace of $X_{cm}$ in the inset of figure~\ref{fig:oneway-interface}\textit{a}. 
% These oscillations are, of course, due to the action of respiratory airflow on the film. (As a test, we reduced the values of $\Pi_\mu$ and $\Pi_\rho$ by a factor of ten and found that the oscillations practically vanish.)

Is the airflow reproduced as accurately by the one-way coupled model? Figure~\ref{fig:oneway-vel} compares time traces of (\textit{a}) the root-mean-square mucus velocity, and (\textit{b}) the radially-averaged axial velocity below the hump, obtained from the fully coupled and the one-way coupled models. We see that though the mucus flow is not entirely reproduced---the small air-induced oscillations are missing---the airflow does appear to be accurately predicted by the one-way coupled model. 

\begin{figure}
\includegraphics[width=1.0\textwidth]{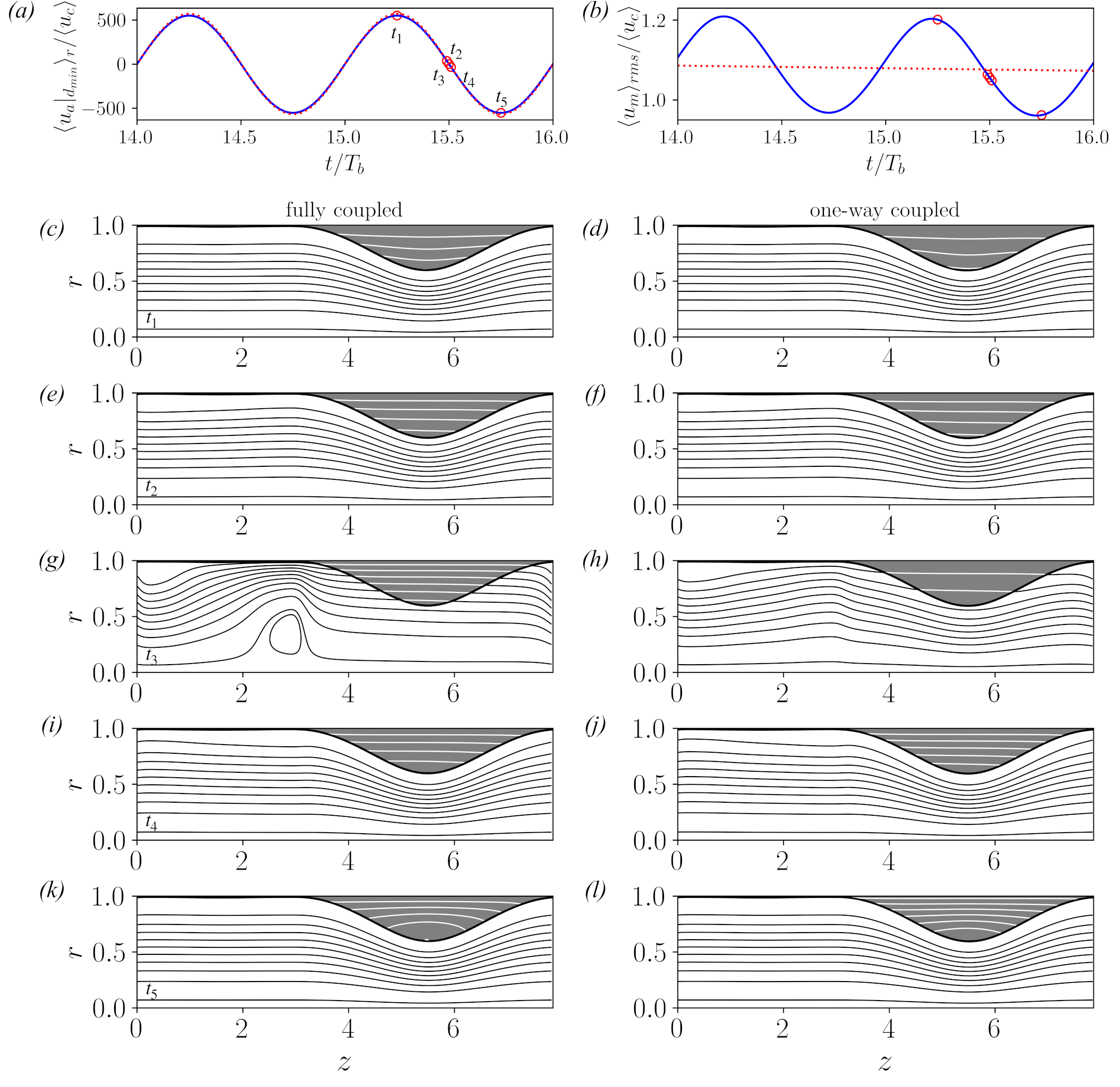}% Here is how to import EPS art
\caption{\label{fig:oneway-streamlines} 
Detailed comparison of the air and mucus flows predicted by the fully-coupled and one-way coupled WRIBL models; the latter ignores the influence of air on the mucus film. Panels (\textit{a}) and (\textit{b}) are temporal zooms of figures~\ref{fig:oneway-vel}(\textit{b})  and \ref{fig:oneway-vel}(\textit{a}), respectively, showing the temporal variation of the mean air velocity (below the hump) and the mean mucus velocity. The line-legend is the same as that in figure~\ref{fig:oneway-vel}. Five time-points are chosen, labelled $t_1$ to $t_5$, corresponding to times when the airflow is opposite to cilia-transport ($t_1$), is in the same direction as cilia-transport ($t_2$), and is reversing ($t_3 - t_5$). The streamlines in the air and mucus at these five times are compared in panels (\textit{c}-\textit{l}) for the fully-coupled (left column, with inset time-labels) and one-way coupled (right column) models.}
\end{figure}

Figure~\ref{fig:oneway-streamlines} compares the predictions of the airflow is more detail. Panels (\textit{a}) and (\textit{b}) zoom into a few oscillations of the long-term air and mucus flow. The remaining panels present snapshots of the instantaneous streamlines, for five selected time points within one breathing cycle, $t_1 - t_5$ in figure~\ref{fig:oneway-streamlines}(\textit{a}). Comparing the streamline plots in the left column (fully coupled model) with those on the right (one-way coupled model), we see that the air streamlines are perfectly captured by the one-way model except for the tiny duration of time between inhalation and exhalation when the airflow is reversing (see figures~\ref{fig:oneway-streamlines}\textit{g} and \ref{fig:oneway-streamlines}\textit{h} corresponding to time $t_3$). For most of the breathing cycle, the airflow is orders of magnitude faster than the mucus flow (compare the y-axis ranges of figures~\ref{fig:oneway-streamlines}\textit{a} and \ref{fig:oneway-streamlines}\textit{b}), and so the air flows past the mucus film as if it were a stationary solid wall. But, when the air velocity is reversing, its magnitude becomes comparable to that of the mucus velocity for a short period of time; this is when the flows in the two fluids are strongly coupled. However, the corresponding time duration is extremely small, as evidenced by the snapshots at times $t_2$ and $t_4$ just before and after $t_3$ (see figure~\ref{fig:oneway-streamlines}\textit{a}). At both these times, the air streamlines again flow around the mucus as if it were stationary, and the one-way coupled model captures the airflow very well (figures~\ref{fig:oneway-streamlines}\textit{e} and \ref{fig:oneway-streamlines}\textit{f} for $t_2$, \ref{fig:oneway-streamlines}\textit{i} and \ref{fig:oneway-streamlines}\textit{j} for $t_4$).

Thus, from the perspective of particle transport and deposition, the one-way coupled model is well-suited to describe the flow of air: Particles will hardly move during the time for which the airflow is inaccurately predicted by the one-way model, because not only is the corresponding time duration very small but also the airflow speed is nearly zero. We will therefore use the one-way coupled WRIBL model to obtain the interface profile and the air flow field for our analysis of particle transport in \S~\ref{sec:particles}. 

% Though not relevant to the present study, the clear impact of respiratory airflow on the flow field in the mucus humps (\textit{e.g.}, contrast figure~\ref{fig:oneway-streamlines}\textit{c} with \ref{fig:oneway-streamlines}\textit{d}) will be important when studying transport within mucus, of particles both passive and active (e.g., swimming bacteria). We comment further on this interesting avenue for future work in the concluding section. 

\section{Extent of mucus-depleted zones on the wall}\label{sec:depleted}

We have seen, in the previous section, that the mucus film ultimately forms an unduloid-like hump that spans a portion of the wall, leaving the rest of the wall without mucus and exposed to particles (figure~\ref{fig:cilia-interface}\textit{a-c}). We now examine how the extent of these mucus-depleted zones varies on increasing the volume or, equivalently, the initial thickness $1-d_0$ of the film. 

% Recall that the upper limit on the volume of the film beyond which unduloid solutions do not exist, i.e. beyond which the airway will be blocked by a liquid bridge, is given by $1.73\pi R^3$ \citep{Dietze2015,everett1972model}. The corresponding cutoff value for the dimensionless initial interface position, $d_0$, is given by solving $(1-d_0^{2}) = 1.73/\Lambda_{RP}$.  

We have performed numerous simulations of randomly-perturbed films, with different initial thicknesses (1-$d_0$). For every value of $d_0$, an ensemble of 8 independent simulations have been performed, starting from random initial perturbations to the flat film:
\begin{equation}
    d(z,0) = d_0 + 10^{-3} \sum_{m=1}^{10} A_m\,\mathrm{cos}(2 \pi m z R/\Lambda_{RP} + \varphi_m),
\end{equation}
where $A_m$ and $\varphi_m$ are uniformly distributed random numbers between $[0,1]$ and $[0, 2 \pi]$ respectively. Each simulation is run for a sufficiently long time so that the film attains its asymptotic profile.  We use the one-way coupled WRIBL model (airflow does not affect the interface evolution, see \S~\ref{sec:air}) and solve only the mucus-film equations \eqref{kbc_mucus}-\eqref{wribl_q_eq_pas}, along with the constant cilia velocity boundary condition $\eqref{cilia_bc_const}$. Cilia transport, though, has no effect on the depleted zones because it does not alter the shape of the interface (see \S~\ref{sec:cilia}). (We have checked that decreasing the cilia velocity by a factor of ten leaves the extent of depleted zones unchanged.)

In our simulations, the depleted zones are occupied by the precursor film. So, we measure the fraction of the wall occupied by depleted zones, $D_f$, by measuring the length of wall along which $1-d<\delta$, where $\delta \approx 0.01$ is the thickness of the precursor film. 

\begin{figure}
\includegraphics[width=1.0\textwidth]{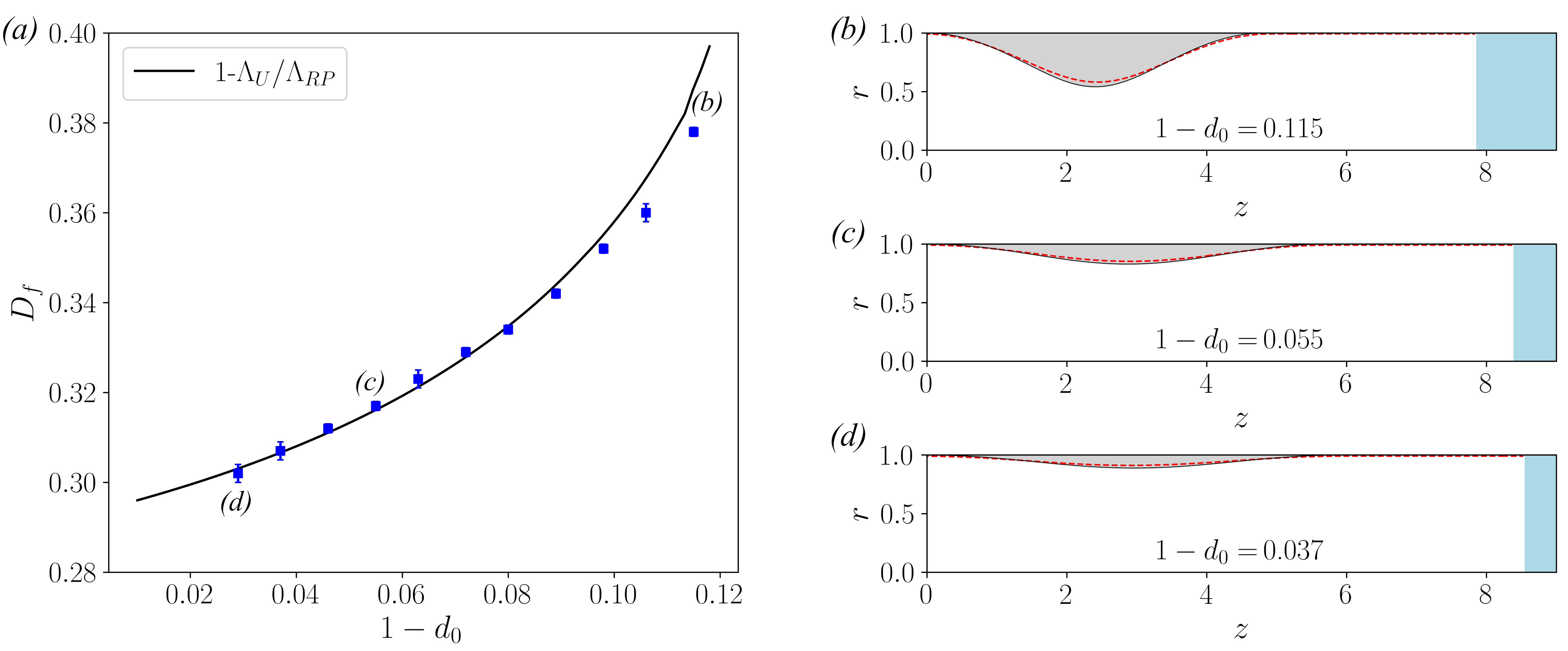}% Here is how to import EPS art
 \caption{\label{fig:dryout_closing} (\textit{a}) Fraction of the wall occupied by mucus-depleted zones. The prediction \eqref{df-eqn} is compared with measurements from randomly-initialized simulations. The markers and bars correspond to the mean and standard deviation of the values measured from the ensemble of eight simulationsf or each value of $d_0$. (\textit{b}-\textit{d}) Shapes of unduloids (shaded in gray) for three values of the initial film thickness, overlaid with the final profile of the film (red-dashed line) from one of the runs in the corresponding ensemble of simulations. The same range of the axial $z$ coordinate is used in these three panels for ease of comparison; the cyan-shaded padding is added to $\Lambda_{RP}/R$ to achieve the same $z$-range.}
\end{figure}

Our results are presented for a range of initial film thicknesses $1-d_0$ in figure~\ref{fig:dryout_closing}(\textit{a}). Here, the markers and bars depict the mean and standard deviation of $D_f$ obtained from the ensemble of simulations for each value of $d_0$. Surprisingly, adding more mucus (increasing $1-d_0$) does not decrease the extent of depleted zones but rather increases it and leaves more of the wall exposed to particles.

This counter-intuitive variation of $D_f$ can be predicted by assuming that all the fluid contained in the initially flat film, of thickness $(1-d_0)R$ and length $\Lambda_{RP}$ (the domain length), is ultimately gathered into an equilibrium-shaped unduloid. This unduloid's axial width $\Lambda_U$ will be less than $\Lambda_{RP}$ (both of which depend on $d_0$), and the difference $\Lambda_{RP}-\Lambda_U$ will be the extent of the mucus-depleted zone. We thus obtain the following prediction for $D_f$:  
\begin{equation}\label{df-eqn}
    D_f \approx 1-\Lambda_U/\Lambda_{RP}.
\end{equation}
$\Lambda_{RP}$, obtained from a linear stability analysis of the WRIBL model, is well approximated by the inviscid result of \citet{Rayleigh1892wavelength}, $\Lambda_{RP} = 2\pi 2^{1/2} d_0 R$ \citep{Dietze2015}. To obtain $\Lambda_U$, we calculate the shape of the unduloid $g(z)$ that (i) has a constant mean-curvature and (ii) encloses the entire volume of the film: 
\begin{multline}\label{unduloid}
             \frac{1}{g}-\frac{(\mathrm{d}_z g)^2}{2g}- \mathrm{d}_{zz}g = K, \;\;g(0) = 1, \;\; \mathrm{d}_z g |_0 = 0, \;\; g(\Lambda_{U}/R) = 1, \\ 2 \pi \int_{0}^{\Lambda_{U}/R} \int_{g(z)}^{1} r dr\, dz =  \pi(1-d_0^2)\Lambda_{RP}/R,
\end{multline}
where $K$ is a constant that is determined by the volume constraint. Note that we have used the $\mathcal{O}(\epsilon^2)$ expression \eqref{curvature} for the mean curvature to be consistent with the WRIBL model. We solve \eqref{unduloid} numerically using a shooting method: We start with a value for $K$ and solve an initial value problem for $g$, stepping forward in $z$ until we satisfy the end boundary condition $g=1$; we thus obtain the unduloid length $\Lambda_U$, after which the unduloid's volume can be calculated. By varying $K$, we obtain unduloids of all volumes less than $1.73 \pi$, beyond which solutions do not exist. These unduloid shapes match quite well with those computed by \citet{everett1972model} using the full expression for the mean curvature; importantly, the dependence of $\Lambda_U$ on the initial thickness of the film (1-$d_0$) is well-approximated (see the \href{https://bighome.iitb.ac.in/index.php/s/6p6PHMqSKYHD6sx}{ supplementary material}). This is not the case if one uses the linearized expression for the curvature (obtained in the limit $1-d_0 \ll 1$), as is done in traditional lubrication models of thin films: the resulting equation is linear in $g$ and yields unduloids with a fixed, $d_0$-independent, width of $2\pi$ \citep{lister2006capillary}. So, in addition to pinchoff dynamics \citep{Dietze2015}, the shape of unduloids is another qualitative feature that necessitates the use of the $\mathcal{O}(\epsilon^2)$ WRIBL model.
% rather than a simpler $\mathcal{O}(\epsilon)$ thin-film model.

The variation of $D_f$ predicted by \eqref{df-eqn} is plotted as a solid-black line in figure~\ref{fig:dryout_closing}(\textit{a}) and seen to agree rather well with the simulations. Importantly, the increase in $D_f$ with the mucus volume can now be understood in terms of the change in the unduloid's shape, which is illustrated in figures~\ref{fig:dryout_closing}(\textit{b-d}). Here, we also overlay the final profile of the film (red-dashed line) from the simulations, on top of the unduloid (gray shading); despite the use of a precursor film, the simulated interface profile agrees rather well with the conceptual ideal of an equilibrium unduloid and a fully-depleted zone. Now, as the film thickness increases, destabilizing capillary forces due to the radial curvature increase in strength relative to stabilizing axial-curvature forces. As a result, the unduloids (and the corresponding humps in the simulations) become deeper, so much so that their length $\Lambda_U$ reduces even as the total volume increases. The wavelength of the fastest-mode $\Lambda_{RP}$ also decreases, but not as much as $\Lambda_U$. The net effect therefore is that a larger portion of the wall is left devoid of mucus.  

The small variation in $D_f$ among the different random simulations (see the small error bars in  figure~\ref{fig:dryout_closing}\textit{a}) confirms our expectation that regardless of the details of the initial perturbation (so long as it triggers the fastest-growing mode of the Rayleigh-Plateau instability) the film ultimately evolves into an unduloid-shaped hump with an adjacent depleted zone. A complete understanding of the role of random perturbations, though, requires an analysis of simulations on long domains that span multiple wavelengths of the fastest growing mode. Our ongoing work in this direction shows that \eqref{df-eqn} remains a good approximation to $D_f$ even on long domains. We comment further on this point in our concluding remarks in \S~\ref{sec:conclusion}.
% \hl{- Error bars are small because only fastest mode is unstable and so results are equiv to single mode, provided linear instability theory holds and determines the final fate - anyway one hump has to form. So random is just a test. }
 %Use toward the end to point to parallel work to test RP: An answer is provided by Rayleigh's paradigm, that the fastest-growing instability mode will dominate and enforce its wavelength onto the pattern of emerging humps \citep{Rayleigh1879}. This principle of wavelength selection works quite well in subcritical interfacial instabilities, like Rayleigh-Plateau and Rayleigh-Taylor, with the exception of special circumstances in which the linear-growth-rate curve (dispersion relation) exhibits multiple peaks \citep{Picardo2017pattern}. The perfect realization of this paradigm in the present problem would have the initially-uniform film of length $L$ organize itself into identical, repeating units of length $\Lambda_{RP}$ (the wavelength of the fastest-growing mode). Indeed, this is the rationale for studying simulations on periodic domains of length $\Lambda_{RP}$, as done in \S~\ref{sec:cilia-air}.
% , and many previous studies as well \citep{lister2006capillary} [ref more].

An important consequence of \eqref{df-eqn} is that the minimum value of $D_f$, attained in the limit of a vanishingly-thin film ($d_0 \rightarrow 1$), is far from zero--- approximately 0.30 as seen from the solid line in figure~\ref{fig:dryout_closing}(\textit{a}). (This limiting value is what is predicted when using the linearized expression for the curvature, as shown in the \href{https://bighome.iitb.ac.in/index.php/s/6p6PHMqSKYHD6sx}{supplementary material}.) So, an appreciable fraction of the wall is always depleted of mucus, leaving it vulnerable to airborne particles. And adding more mucus only worsens the situation by producing deeper but shorter humps that expose even more of the wall. Does this increased exposure imply that more particles will deposit on the wall when more mucus is present, or will the deeper humps be able to trap more particles? We directly address this question, for particles spanning a wide range of sizes, in the next section. 

\section{Particle transport and deposition}\label{sec:particles}

We consider two different scenarios in this study of particle deposition, which correspond to two extremes of inter-breath particle-turnover. In the first scenario, the same particles persist within the airway throughout the simulation, which spans 30 breathing cycles (30 seconds). In the second scenario, we assume that all particles are replaced by fresh ones after each breath, due to mixing between tidal air and reserve air or between neighbouring airways (\citeauthor{airwaymixing1959} \citeyear{airwaymixing1959}; \citeauthor{Wang2005book} \citeyear{Wang2005book}, chap 5).  

A number concentration of 300 \si{\mu g.m^{-3}}, which is typical for $\mathrm{PM}_{2.5}$ \citep{epa}, translates to about 16 micron-sized particles in the airway under consideration. So, in the first scenario, we introduce 16 particles into the periodic airway-domain, at random locations, and track their motion for 30 seconds (while removing particles that deposit on the film). We then perform an ensemble of 125 such simulations, each starting from a different random initial-distribution of particles. We thus obtain good statistics for deposition results, such as $\phi(t)$, the fraction of particles that have deposited (on the wall or mucus) at any time during the simulation. Now, because the particles do not interact with each other in the simulation, the mean value of $\phi$ obtained from the ensemble of $125$ simulations would be identical to the value obtained from a single simulation with $16*125 = 2000$ particles. Hence, the chosen number concentration, \textit{i.e.}, the number of particles introduced into the airway in each simulation, will only affect the variance in the deposition fraction and not the ensemble-averaged mean results.

In the second scenario, we again introduce 16 particles at random locations in the airway and track them for one breathing cycle; we then remove all non-trapped particles and introduce a fresh set of 16 particles. This process is repeated for 30 breathing cycles (30 second of respiration). We then perform an ensemble of 60 such simulations. Of course, because the particles are replaced after each cycle, this calculation is equivalent to running an ensemble of 1800 simulations, each for one breathing cycle. 

% The number of particles in a single simulation does matter, however, for the second scenario because the particles are replaced after every breath by new particles at ne random positions. 
From the perspective of particle transport, the key difference between the two scenarios is that only in the first scenario will the distribution of particles have sufficient time to respond to the non-homogeneous nature of the flow field. For example, the curved streamlines of the airflow near the mucus hump will induce a slight centrifugal-drift of weakly inertial particles towards the centreline, which in the first scenario will add up from breath to breath to produce an accumulation of particles near the centreline. Such accumulative effects will not occur in the second scenario, wherein complete turnover of particles takes place after each breath.

While discussing our results we shall distinguish between deposition on the mucus humps (mucus entrapment) and deposition on the mucus-depleted zones (wall deposition). In the latter case, the particles deposit not directly on the epithelial cells of the airway wall, but rather on the PCL layer which our model ignores. Nevertheless, we shall, for simplicity, use the term wall-deposition because these particles are much more likely to reach the wall through the thin, low-viscosity, PCL layer than through the combination of the thick, viscous, mucus hump followed by the PCL layer. 

In \S~\ref{sec:air}, we had noted that the air flow was orders of magnitude faster than the cilia-induced translation of the interface and so we had anticipated that ciliary transport will not affect particle deposition. We have checked this and indeed found that our deposition results remain unaffected by decreasing $\langle u_c \rangle$ by a factor of ten. So, though ciliary transport is present in the particle-transport simulations, it does not contribute to deposition. 

\begin{figure}
\includegraphics[width=1\textwidth]{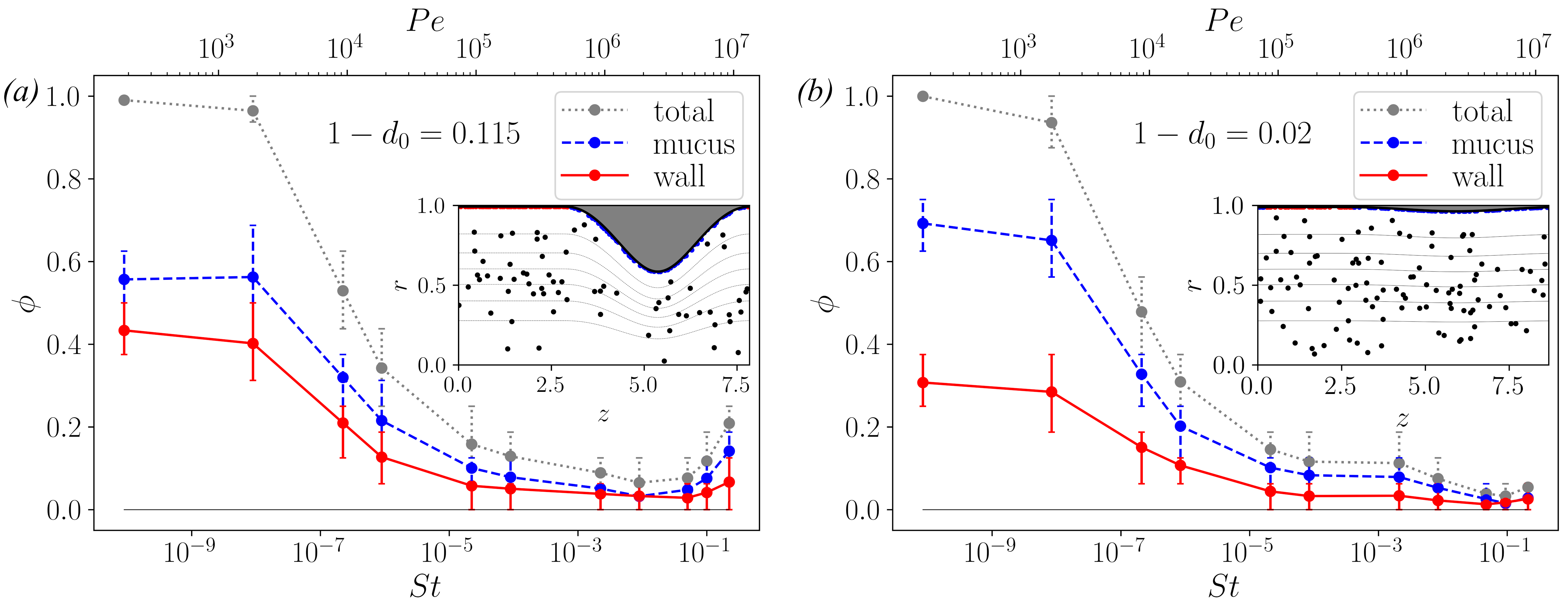}% Here is how to import EPS art
\caption{\label{fig:trapping_30s}  Deposition fraction $\phi$ as a function of particle size, after 30 breathing cycles (scenario one), in an airway with (\textit{a}) a relatively thick film, $1-d_0=0.115$, and (\textit{b}) a very thin film, $1-d_0 = 0.02$. The particle size is represented by the corresponding values of $St$ (bottom axis) and $\Pen$ (top axis). Snapshots of the two films, in their long-term asymptotic states, are shown as insets in the corresponding panels. For each particle size, the marker represents the mean of the distribution of $\phi$, obtained from the ensemble of 125 simulations; the bar corresponds to the interquartile range which is indicative of the spread of the distribution. The fraction of particles depositing on the mucus hump, the depleted-zone of the wall, and on either surface are shown separately (see the legend).}
\end{figure}

Let us begin with the first scenario and examine how the deposition fraction $\phi$ 
% (fraction of particles introduced into the airway that are deposited at time $t$) 
depends on the particle size. Results obtained after 30 breathing cycles (30 seconds) are presented in figure~\ref{fig:trapping_30s}(\textit{a}) for a relatively thick film. Three curves of $\phi$ are shown, corresponding to the fraction of particles deposited on the mucus hump, on the airway wall (depleted zone), and on either surface (total deposition). The particle size is represented by $\Pen$ and $St$ (see \eqref{Pe-St} and table~\ref{table:particles}); recall that $\Pen$ represents the importance of convection relative to Brownian-diffusion, while $St$ indicates the importance of particle inertia. Both $\Pen$ and $St$ increase with size, so that the particle motion changes from diffusive to tracer-like to inertial. This varying nature of particle motion may be appreciated by comparing supplementary \href{https://bighome.iitb.ac.in/index.php/s/o4oTCqQpzXjnCxB}{movie 1} with \href{https://bighome.iitb.ac.in/index.php/s/r48Gi7cTk9ZjM3q}{movie 2} and \href{https://bighome.iitb.ac.in/index.php/s/yPTomcKgbmnY2MG}{movie 3}, which illustrate the motion of particles with $\Pen = 2 \times 10^4$ ($St = 9 \times 10^{-7}$), $\Pen =  9 \times 10^5$ ($St = 2 \times 10^{-3}$), and $\Pen =  9 \times 10^6$ ($St = 2 \times 10^{-1}$), respectively. (These movies use a large number of particles to better visualize the qualitative features of their motion.)

Figure~\ref{fig:trapping_30s}(\textit{a}) shows that the total deposition initially decreases with particle size, as the strength of Brownian forces reduce. However, there is an increase of deposition for the larger, inertial particles. That this significant increase is associated with the presence of a deep mucus hump is demonstrated by comparing figure~\ref{fig:trapping_30s}(\textit{a}) with figure~\ref{fig:trapping_30s}(\textit{b}). In the latter, which corresponds to a much thinner film with a shallow hump, the deposition fraction remains very small even for inertial particles. 

% \begin{table}
% \begin{center}
% \phantom{~}\noindent
%   \begin{tabular}{p{0.1\textwidth}>{\centering}p{0.1\textwidth}>{\centering}p{0.2\textwidth}>{\centering}p{0.2\textwidth}>{\centering}p{0.1\textwidth}>{\centering}p{0.1\textwidth}>{\centering\arraybackslash}p{0.1\textwidth}}

%   $Data$ $set$&$d_0$&$volume fraction$ & $\tau_{mucus}/\tau_{wall}$ & $A_f^{mucus} $ & $A_f^{wall}$ & $L_D$ \\

%  $1.$&$0.885$&$ 1.70$ & $0.34$& $0.55$ & $0.45$ & $0.397$ \\  
%  $2.$&$0.98$&$ 0.34$ & $0.95$& $0.67$ & $0.33$ & $0.306$ \\
% \end{tabular}
% \caption{particles}
% \label{table:particles}
% \end{center}
% \end{table}

The non-monotonic variation of total deposition with particle size, seen in figure~\ref{fig:trapping_30s}(\textit{a}), is a well known feature of experimental measurements of deposition in the lungs \citep{morawska2022physics}. However, the increase for inertial particles is much stronger in entire-lung measurements because the effects of particle inertia are very strong in the turbulent flows of the upper airways, as well as in the secondary circulatory flows at airway bifurcations~\citep{guha2008transport}. Here, we see that a non-monotonic variation of deposition, albeit weaker, arises even in a single airway provided the mucus film is thick enough to form prominent humps.

\begin{figure}
% \begin{center}
    \includegraphics[width=1\textwidth]{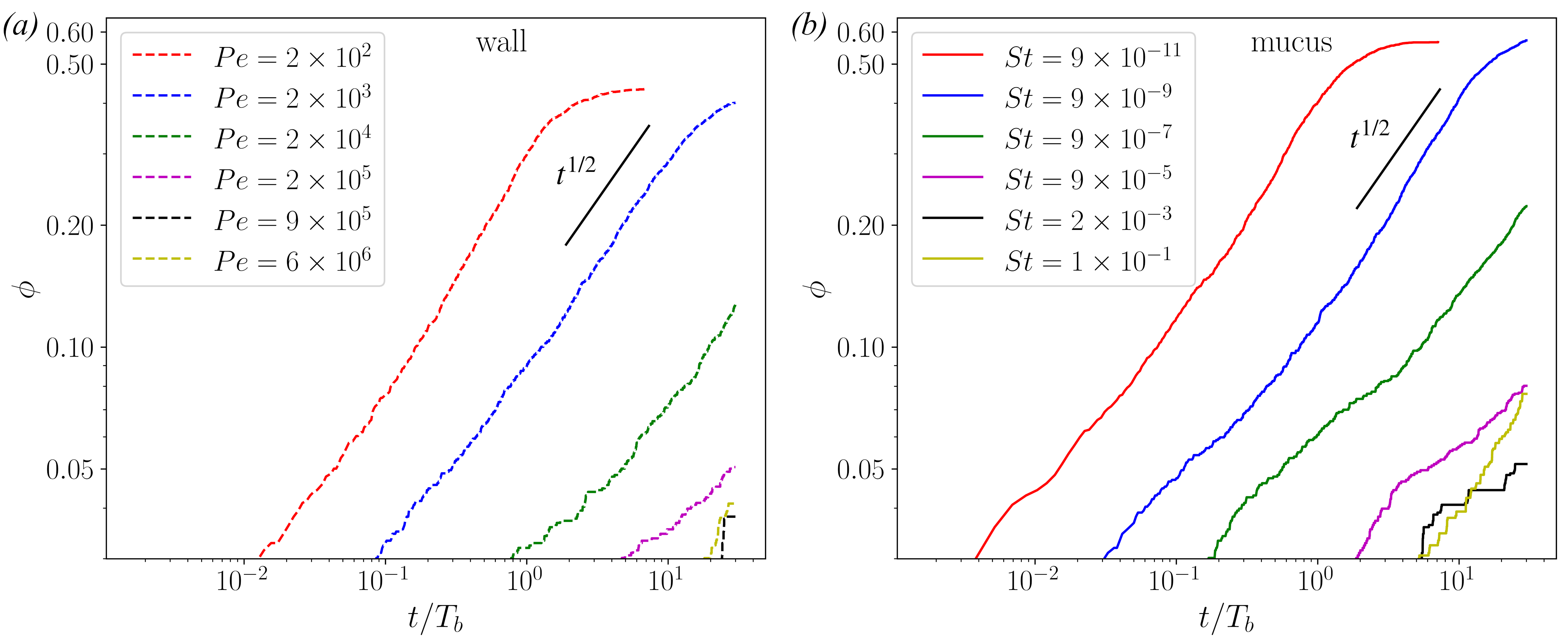}% Here is how to import EPS art
\caption{\label{fig:phi_t} Evolution of the ensemble-averaged deposition fraction on (\textit{a}) the depleted-zones of the wall and (\textit{b}) the mucus hump. Results are shown for six different particles sizes; for convenience, both the $\Pen$ and $St$ values of these particles are mentioned in the legend, which is split across the two panels. The solid-black line corresponds to a growth of $t^{1/2}$.}
% \end{center}
\end{figure}

The variation of the deposition fraction with time, for different particle-sizes, is presented in figure~\ref{fig:phi_t}. The small particles, whose deposition is driven by Brownian motion, naturally exhibit an approximate $t^{1/2}$ diffusive growth of $\phi$. Given sufficient time, all these particles will be deposited on either the wall or the mucus humps. The effects of inertia are first seen for $St = 2 \times 10^{-3}$, for which $\phi$ saturates to a low value (figure~\ref{fig:phi_t}\textit{b}). As shown below, the cessation of further deposition is a result of gradual centrifugal migration towards the centreline, owing to the curved air streamlines below the mucus hump. As these inertial forces become stronger, the particles veer-off from streamlines more violently and collide with the mucus hump; hence the increase in deposition for the largest particle with $St = 2 \times 10^{-1}$ (figure~\ref{fig:phi_t}\textit{b}).

Returning to figure~\ref{fig:trapping_30s}(\textit{a}), let us examine the difference between the deposition on the mucus hump (blue dashed line) and on the exposed wall (red solid line). For the small particles, we see that though a higher fraction deposit on the mucus hump, a significant fraction does reach the wall. The difference between the deposition fractions mirrors the difference in the surface areas of the mucus hump and the exposed wall (55\% and 45\%, respectively, of the total surface area exposed to air). This correspondence is a sign of the diffusive nature of deposition of the small particles (see in particular the values for the smallest particle with $\Pen = 2 \times 10^2$). 
% For larger particles, with higher $\Pen$, the deposition on both the mucus and the wall decreases as the effects of Brownian motion weaken. 
% we see that the difference in deposition shrinks. This is a consequence of convection by the airflow, which is faster below the hump (where the cross-sectional area for flow is the least). Particles thus spend more time near the depleted zone of the wall (about 3 times longer than below the hump), which offsets the effect of the higher surface area of the hump. 
The deposition of inertial particles on the hump and the wall is governed by other physical effects and will be discussed later in this section.

\begin{figure}
\includegraphics[width=1\textwidth]{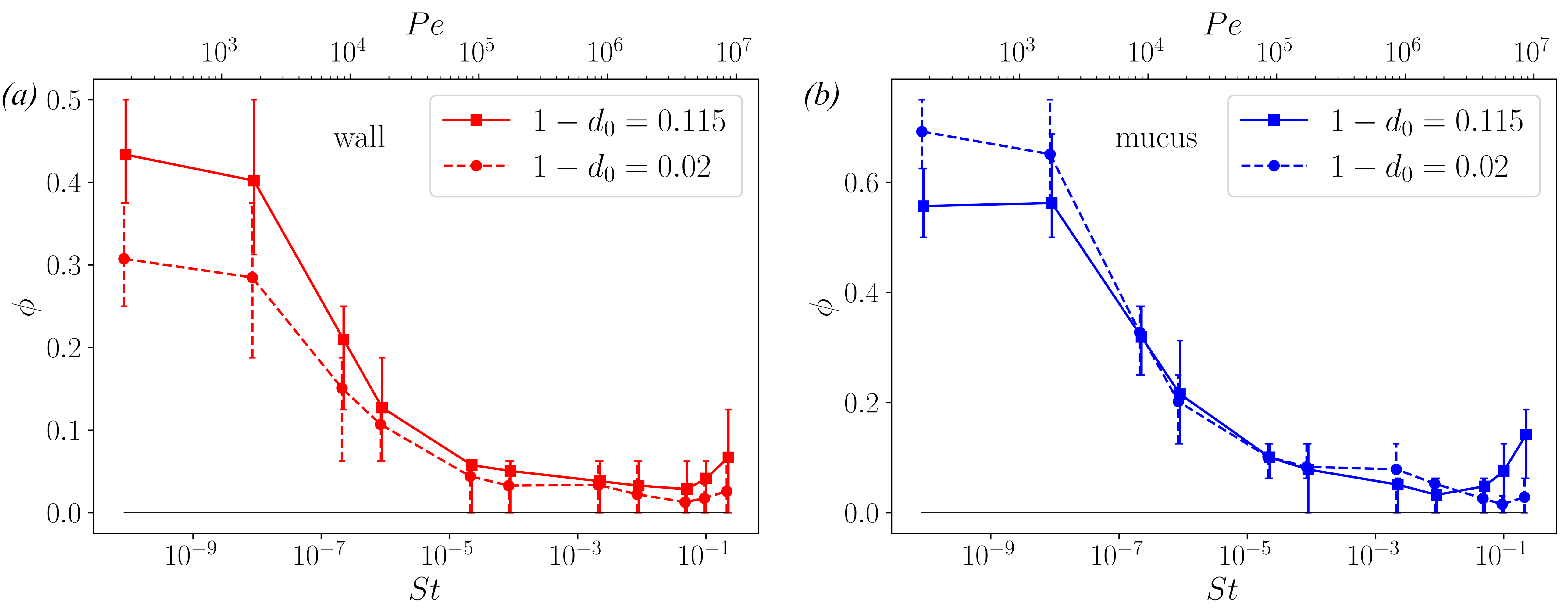}% Here is how to import EPS art
\caption{\label{fig:trapping-d0} 
Effect of the initial film thickness (mucus volume) on the deposition fraction. The results from figures~\ref{fig:trapping_30s}(\textit{a-b}) are overlapped for deposition on (\textit{a}) the depleted-zone of the wall, and (\textit{b}) the mucus hump.}
\end{figure}

Next, let us examine the effect of the mucus volume fraction with the aid of figure~\ref{fig:trapping-d0}, which overlaps the results for the two film thickness shown previously in figures~\ref{fig:trapping_30s}(\textit{a}-\textit{b}). We have already seen that thicker films produce larger mucus-depleted zones (see figure~\ref{fig:dryout_closing}\textit{a}). One would expect the widening of depleted zones to allow more small Brownian particles to deposit on the wall when the film is thicker. This is exactly what we observe in figure~\ref{fig:trapping-d0}(\textit{a}). Consequently, a smaller fraction of small particles is entrapped by the mucus humps in the case of the thicker film (figure~\ref{fig:trapping-d0}\textit{b}). These differences between the two films are less for the larger tracer-like particles, which diffuse more slowly. However, deposition of the largest, inertial particles is again markedly higher for the thicker film---especially on the mucus hump---owing to inertial effects. 

\begin{figure}
\includegraphics[width=1\textwidth]{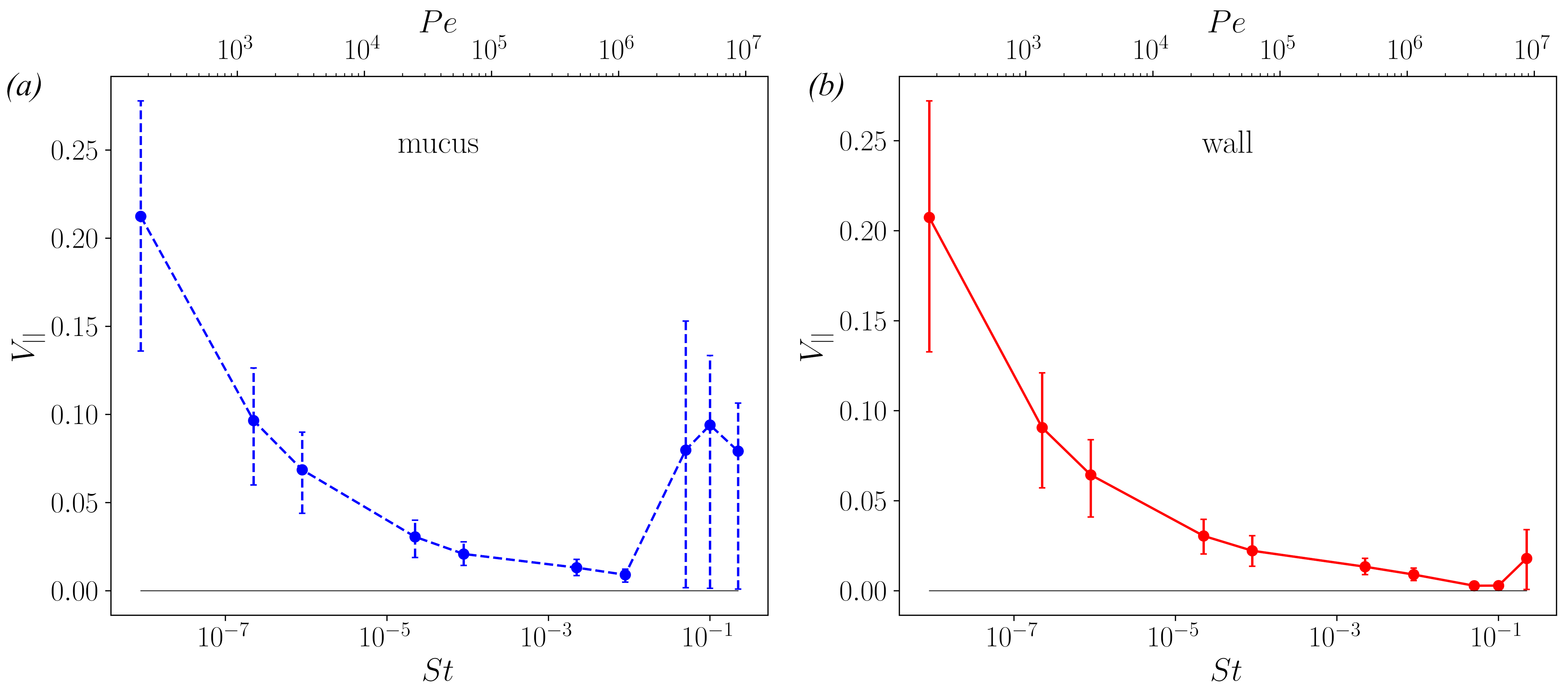}% Here is how to import EPS art
\caption{\label{fig:rvel} Collision velocity $V_{||}$ of particles with the depositing surface: (\textit{a}) mucus hump, (\textit{b}) depleted-zone of the wall. The marker and the bar represents the mean and the inter-quartile-range (which measures the variation about the mean) of the distribution of $V_{||}$ for each particle size. These results correspond to the case of the relatively thick film in figure~\ref{fig:trapping_30s}(\textit{a}). }
% The mean of the distribution of $V_{||}$, for each particle size, is represented by the marker, while the inter-quartile-range (which measures the spread of the distribution) is depicted by the bar.}
\end{figure}

Inertia causes the large particles to deviate from the air streamlines and collide with the wall and the mucus hump.
% The mechanism that produces higher mucus entrapment of inertial particles (figure~\ref{fig:trapping-d0}\textit{a}), for the thicker film, is different from the mechanism underlying the higher wall deposition of inertial particles (figure~\ref{fig:trapping-d0}\textit{b}). 
When heavy particles approach the mucus hump, their inertia prevents them from going around the hump along with the curved streamlines. Instead, they continue moving straight and collide with the mucus hump. Evidence for such inertial impaction \citep{guha2008transport} is provided by the collision velocity, $V_{||}$, which is the relative velocity between the particle and the deposition site (mucus hump or wall) at the time of deposition, projected along the approach direction. Figure~\ref{fig:rvel} presents the mean collision velocity of particles with the mucus hump (panel \textit{a}) and the wall (panel \textit{b}) for the case of the thick film (figure~\ref{fig:trapping_30s}\textit{a}); here, the bars represent the inter-quartile range which is indicative of the spread of the distribution of $V_{||}$. The smallest diffusive particles have a large $V_{||}$ because of the dominance of fluctuating Brownian forces. The tracer-like particles have a very low $V_{||}$, as expected for particles whose velocity is almost entirely aligned with the air streamlines which are tangential to the mucus hump and the wall. The relatively large, inertial particles though show a marked increase in $V_{||}$, especially for collisions with the mucus hump which protrudes into the airway (figure~\ref{fig:rvel}\textit{a}); as anticipated, inertial particles deviate from the streamlines and collide with the mucus hump.  

The slight decrease in $V_{||}$ for the collisions of the largest particle with the hump (compare the second-largest with the largest particle in figure~\ref{fig:rvel}\textit{a}) signifies the growing influence of the particle interception effect \citep{guha2008transport}: Because of its increasing radius, the centre of the particle need not deviate as strongly from the airflow streamlines in order for the particle surface to make contact with the mucus film. Thus, deposition will include more low-velocity collisions as the particle size increases. 

% Note that a higher $V_{||}$ translates to a higher deposition rate, since the latter is a surface integral of $V_{||}$ times the surface area element times the number concentration near the surface. Thus, $V_{||}$ in figure~\ref{fig:rvel} reflects the non-monotonic trend of $\phi$ in figure~\ref{fig:trapping_30s}(\textit{a}).

The collision velocity of inertial particles with the wall is much smaller that that with the hump, even for the largest particle (compare figures~\ref{fig:rvel}\textit{a-b}).
Deposition on the wall, though, is aided by the
% Yet, the corresponding deposition fractions are nearly the same (figure~\ref{fig:trapping_30s}\textit{a}). The explanation lies in the 
non-uniform number concentration of inertial particles in the airway: a subtle interplay of varying airflow speed and streamline-curvature leads to an elevated number concentration near the depleted zone on the wall.

\begin{figure}
\includegraphics[width=1\textwidth]{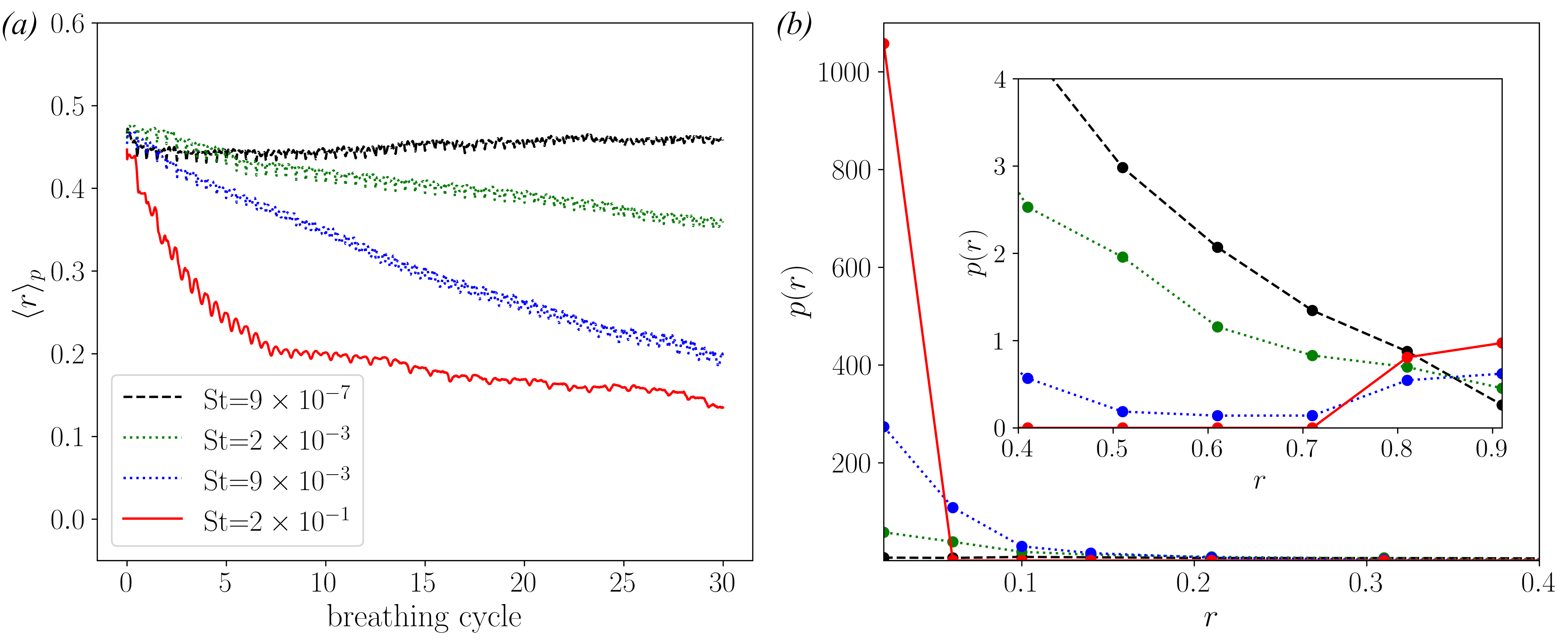}% Here is how to import EPS art
\caption{\label{fig:part-dist} Non-homogeneous distribution of particles in the airway, for the case of the relatively thick film (see figure~\ref{fig:trapping_30s}\textit{a}). (\textit{a}) Evolution of the mean radial location of particles (non-deposited) in the airway; the migration of inertial particles ($St = 0.009,\,0.2$) towards the centreline is clearly visible. (\textit{b}) The probability distribution function of the radial location of particles, integrated over time (data from $t = 0$ to $t = 30T_b$ is used to construct the distribution). The distribution near the centreline is shown in the main panel, while that near the wall is show in the inset. }
\end{figure}

The spatial distribution of particles is examined in figure \ref{fig:part-dist}. The mean radial location of all non-trapped particles, $\langle r \rangle_p$, is plotted as a function of time in figure \ref{fig:part-dist}(\textit{a}). The gradual migration towards the centreline, $r = 0$, is apparent here for the relatively-large inertial particles; such migration does not occur for the small diffusive particles. The small oscillations in $\langle r \rangle_p$, present for all particles, are a result of the axisymmetric geometry and the converging-diverging nature of the air streamlines, which causes a small convergence toward the centreline, followed by a corresponding divergence, each time particles are carried past the hump. These oscillations do not lead to a net convergence, though, and the preferential concentration near the centreline is caused by gradual inertial centrifugation. 

The probability distribution function (pdf) of the radial location of non-trapped particles, integrated over time, is show in figure \ref{fig:part-dist}(\textit{b}). The main panel reinforces the message of figure \ref{fig:part-dist}(\textit{a}), that small particles are uniformly distributed, while inertial particles accumulate near the centreline. The inset, though, reveals a surprising accumulation of inertial particles near the depleted zone of the wall (there is no accumulation below the hump which extends up to $r \approx 0.6$). Supplementary \href{https://bighome.iitb.ac.in/index.php/s/yPTomcKgbmnY2MG}{movie 3} shows how heavy particles that begin near the depleted zone tend to remain there (though the majority of particles, which at some time pass by the hump, are centrifuged towards the centreline). This is due, in part, to the low air-velocity near the depleted zone (which has the highest cross-sectional flow area). Over a breathing cycle, the particles travel only a short way past the edge of the hump before returning to the depletion zone, and hence they do not pass the bottom of the hump where the centrifugation towards the centreline acts most strongly. On the contrary, because the curvature of the streamlines near the edge of the depletion zone (where it meets the hump) is opposite to that of the streamlines below the hump, the inertial particle are weakly pushed towards the edge of the depletion zone. 

% The importance of inertia for this near-wall accumulation is clear in the inset of \ref{fig:part-dist}(\textit{b})

To summarize the results of the first scenario, there is a non-monotonic variation of deposition with particle size because small particles deposit due to Brownian forces and large particles deposit due to inertial effects. The latter are rather subtle: strong centrifugal forces near mucus humps produce high-velocity collisions, whereas weak centrifugal forces near mucus-depleted zones of the wall cause particles to accumulate and then gradually deposit there.

\begin{figure}
\includegraphics[width=1\textwidth]{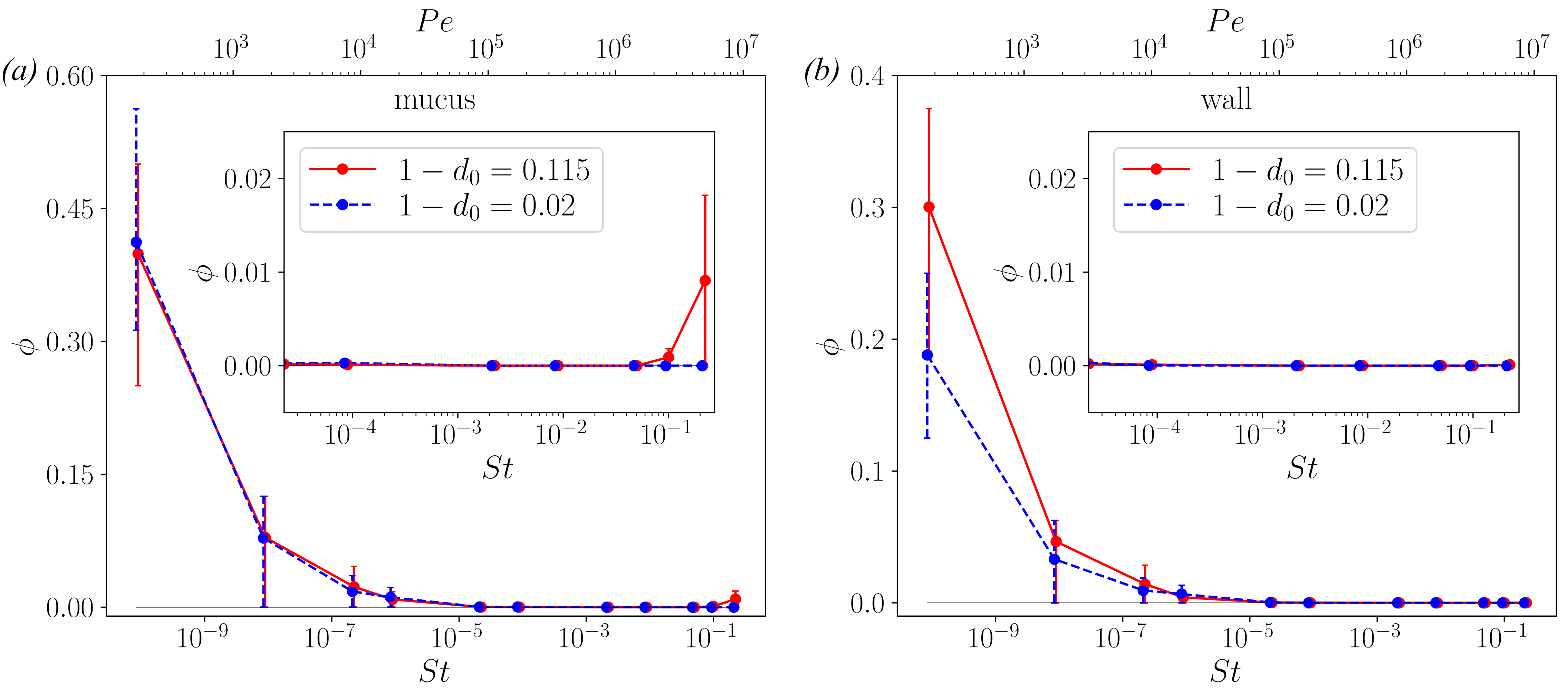}% Here is how to import EPS art
\caption{\label{fig:trapping_turnover} 
Variation of the deposition fraction $\phi$ with particle size, for the second scenario in which particles are replaced after each breath. Results for the thick and thin film are compared for deposition on (\textit{a}) the mucus hump and (\textit{b}) the wall. The insets are zooms that show the variation of $\phi$ for the large particles. The markers and the bars represent the mean and the inter-quartile-range of the deposition fraction measured from an ensemble of 60 simulations, each run for 30 breathing cycles (a total of 1800 breaths between which particles are entirely replaced).}
% See: \url{https://stackoverflow.com/questions/22995797/can-matplotlib-errorbars-have-a-linestyle-set}
\end{figure}

Let us now turn to the second scenario. The deposition fraction here will be smaller than in the first scenario because the particles have just one breathing cycle (instead of thirty) in which to deposit before they are replaced by new particles. At the same time, the airway in the second scenario is exposed to far more particles, in total. So, while interpreting the magnitude of $\phi$, one should bear in mind that the number of deposited particles in the first and second scenarios is given by $16 \phi$ and $(16\times30) \phi$, respectively. In the following discussion of the second scenario, we shall examine how $\phi$ changes with the particle size, in light of the corresponding variation in the first scenario.

Because the particles are replaced very frequently, the inertial particles will not have time to accumulate near the centreline or near the wall, as they did in scenario one (figure~\ref{fig:part-dist}\textit{b}). Therefore, while particle inertia should still promote deposition on the mucus hump due to centrifugation and high-velocity collisions, there should not be any wall-deposition due to inertia because particles will not accumulate near the wall in scenario two. This is exactly what we see in figure~\ref{fig:trapping_turnover}, which compares the results for deposition on the mucus hump (panel \textit{a}) and on the mucus-depleted portion of the wall (panel \textit{b}), for the two cases of a thick and a thin film (see the legend). More inertial particles are trapped by the thicker film with the deeper mucus hump (see the inset of figure~\ref{fig:trapping_turnover}\textit{a} and note the significant relative increase of $\phi$ for the largest particle). As anticipated, there is no such increase in the case of wall-deposition of inertial particles (see the inset of figure~\ref{fig:trapping_turnover}\textit{b}). 
For small diffusive particles, we see that, just as in scenario one, the thicker mucus film allows more particles to deposit on the wall because of the presence of larger mucus-depleted zones (figure~\ref{fig:trapping_turnover}\textit{b}).
% correspondingly, the thicker film traps fewer small particles in the mucus humps (figure~\ref{fig:trapping_turnover}\textit{a}). 

The key takeaway from the results of the second scenario is that particle turnover counter-acts the effects of inertia-driven preferential accumulation. So, if a substantial fraction of particles in an airway are replaced after each breath (thanks to mixing between airways or between tidal and reserve air volumes), then a higher mucus volume will not cause more wall-deposition of heavy inertial particles. Taken together, the two scenarios suggest that the primary effects of increasing the mucus volume fraction are more mucus-entrapment of large inertial particles and more wall-deposition of small diffusive particles. Which of these opposing outcomes is detrimental or beneficial to human health depends on the nature of the particles---harmful allergens and pathogens, or therapeutic drugs.

\section{Concluding remarks}\label{sec:conclusion}

We have studied the transport of aerosols in a ciliated mucus-bearing airway with the aim of understanding how the distribution of the mucus film, and the airflow around it, affects particle deposition. Considering a typical mid-generation airway, our simulations and analyses have shown that the mucus distribution is unaffected by airflow, which is too weak (under normal breathing conditions) to compete with the capillary forces that determine the final shape of the mucus interface. Ciliary transport also does not alter the interface profile but simply translates it. The speed of translation is orders-of-magnitude slower than the velocity of airborne particles and so ciliary transport has no impact on the nature or rate of deposition.

While a thick film will pinch-off (and form a liquid bridge that occludes the airway), a sufficiently thin film will remain open, organizing itself into unduloid-shaped annular humps separated by depleted zones. Focusing on open airways, we have investigated the transport of air-borne particles spanning a wide range of sizes. We have seen that the distribution of mucus into humps and depleted zones significantly impacts particle deposition, and in a manner than depends on the particle size. Our counterintuitive finding, that thicker mucus films produce larger depleted zones, results in more small (diffusive) particles depositing on the wall, as the amount of mucus is increased. For large (inertial) particles, however, a thicker film aids in mucus entrapment by producing deeper humps that intercept more particles. 

The analysis in this work has been based entirely on a periodic domain of length equal to the wavelength of the fastest-growing Rayleigh-Plateau mode. Typical mid-generation airways have lengths that are up to two to four times longer (table~\ref{table:properties}). So, our results will apply to long airways if the film organizes itself into a repeating pattern with wavelength roughly equal to that of the fastest mode. This dominance of the fastest-growing instability mode, first hypothesized by \citet{Rayleigh1879}, is typically true in problems of interfacial pattern formation like Rayleigh-Plateau and Rayleigh-Taylor, with the exception of special circumstances in which the linear-growth-rate curve (dispersion relation) exhibits multiple peaks \citep{Picardo2017pattern}. For the present problem, we expect Rayleigh's paradigm to hold. We have tested this expectation by comparing the results of long domain simulations, spanning multiple fastest-mode wavelengths, against those of the short, single-wavelength domain. With regard to the extent of mucus-depleted zones, our results show that the analysis of \S~\ref{sec:depleted} works reasonably well even on long domains \citep{Hazra2025}.

Several interesting, physiologically-relevant extensions to our study are possible. For example, one could incorporate allergen-induced mucus secretion~\citep{allergen2015}, such that the deposition of particles on the wall triggers a release of mucus~\citep{Silberberg}. Such a model would shed light on the rapid blockage of airways by mucus plugs (liquid bridges), as occurs in cases of sudden-onset fatal asthma~\citep{hays2003role,rogers2004airway}. The model can also be extended to describe the dynamics of the plugs so formed, by following the recent work of \citet{Dietze2024plugs}, which represents a major advance in thin-film modelling. By detailed comparisons with volume-of-fluid based direct-numerical-simulations, \citet{Dietze2024plugs} shows that plugs can be faithfully described by an extended form of the WRIBL model, which was first proposed by \citet{Dietze2020occlusion}. This model adds a source term which applies a disjoining-pressure at the centreline to stabilize the closing interface in the form of a pseudo-plug.

% Another avenue for future study is to investigate particle entrapment under conditions of stronger airflow, as occurs during post-exercise, open-mouth breathing \citep{tsuda2013particle}.waves on mucus and \citet{kim1985in-vivo,kim1985in-vitro}\\

Our study has shown that depositing particles can avoid mucus due to the ever-present mucus-depleted zones along the wall. While such direct deposition of particles certainly poses a threat to the health of the airway, even the particles that are intercepted by mucus humps could still find their way, through the mucus, to the wall. While studying in-mucus transport, it will be important to consider 
% the alteration to the mucus flow produced by oscillatory air-drag (as shown by the comparison between the one-way coupled and fully coupled WRIBL models in \S~\ref{sec:air}). Ciliary transport, which is too slow to affect the deposition of particles from the air, will be an important factor when studying particle transport inside mucus. Indeed, \citet{Dietze2023mucociliary} show that ciliary forcing alters the mucus flow field in a manner that depends on the amount of lubrication provided by the PCL layer (modelled as a partial slip in the coarse-grained cilia boundary condition) and on 
the non-Newtonian rheology of mucus, which affects particle transport by altering not only the diffusive motion of particles \citep{hill2014biophysical} but also the active motility of microorganisms like bacteria \citep{Laugabook,Pednekar22}.

% This study has shown that depositing particles can avoid mucus due to the ever-present mucus-depleted zones along the wall. While such direct deposition of particles certainly poses a threat to the health of the airway, even the particles that are intercepted by mucus humps could still find their way, through the mucus, to the wall. While studying in-mucus transport, it will be important to consider the alteration to the mucus flow produced by oscillatory air-drag (as shown by the comparison between the one-way coupled and fully coupled WRIBL models in \S~\ref{sec:air}). Ciliary transport, which is too slow to affect the deposition of particles from the air, will be an important factor when studying particle transport inside mucus. Indeed, \citet{Dietze2023mucociliary} show that ciliary forcing alters the mucus flow field in a manner that depends on the amount of lubrication provided by the PCL layer (modelled as a partial slip in the coarse-grained cilia boundary condition) and on the viscoelasticity of mucus. The non-Newtonian rheology of mucus is important, even in the absence of flow, because it affects particle transport by altering not only the diffusive motion of particles \citep{hill2014biophysical} but also the active motility of microorganisms like bacteria \citep{Laugabook,Pednekar22}.

\vspace{.5\baselineskip}
\noindent\small{\textbf{Acknowledgements.} {J.R.P. acknowledges his Associateship with the International Centre for Theoretical Sciences (ICTS), Tata Institute of Fundamental Research, Bangalore,
India. In particular, he is very grateful for the warm hospitality he received at ICTS during the writing of this paper. The authors thank the National PARAM Supercomputing Facility \textit{PARAM SIDDHI-AI} at CDAC, Pune for computing resources; simulations were also performed on the IIT Bombay workstations \textit{Gandalf} (procured through DST-SERB grant SRG/2021/001185), and \textit{Faramir} and \textit{Aragorn} (procured through the IIT-B grant RD/0519-IRCCSH0-021).}

% \backsection[Acknowledgements]{J.R.P. acknowledges his Associateship with the International Centre for Theoretical Sciences (ICTS), Tata Institute of Fundamental Research, Bangalore,
% India. In particular, he is very grateful for the warm hospitality he received at ICTS during the writing of this paper. The authors thank the National PARAM Supercomputing Facility \textit{PARAM SIDDHI-AI} at CDAC, Pune for computing resources; simulations were also performed on the IIT Bombay workstations \textit{Gandalf} (procured through DST-SERB grant SRG/2021/001185), and \textit{Faramir} and \textit{Aragorn} (procured through the IIT-B grant RD/0519-IRCCSH0-021).}

\vspace{.5\baselineskip}
\noindent\small{\textbf{Funding.} {This work was supported by
DST-SERB (J.R.P., grant no. SRG/2021/001185), the Indo–French Centre for the Promotion of Advanced Scientific Research (IFCPAR / CEFIPRA) (J.R.P. project no. 6704-A), and IRCC, IIT Bombay (S.H., Ph.D. fellowship; J.R.P., grant no. RD/0519-IRCCSH0-021).
}

% \backsection[Funding]{This work was supported by
% DST-SERB (J.R.P., grant no. SRG/2021/001185), the Indo–French Centre for the Promotion of Advanced Scientific Research (IFCPAR / CEFIPRA) (J.R.P. project no. 6704-A), and IRCC, IIT Bombay (S.H., Ph.D. fellowship; J.R.P., grant no. RD/0519-IRCCSH0-021).
% }

% \backsection[Declaration of interests]{The authors report no conflict of interest.}

% \backsection[Data availability statement]{The data that support the findings of this study are available from the authors on reasonable request.}

\vspace{.5\baselineskip}
\noindent\small{\textbf{Author ORCHID.} {S.~Hazra, https://orcid.org/0009-0000-3728-5888; J.~R.~Picardo, https://orcid.org/0000-0002-9227-5516}}
% \backsection[Author ORCID]{S.~Hazra, https://orcid.org/0009-0000-3728-5888; J.~R.~Picardo, https://orcid.org/0000-0002-9227-5516}

\appendix

\begin{figure}
\begin{center}
\includegraphics[width=.5\textwidth]{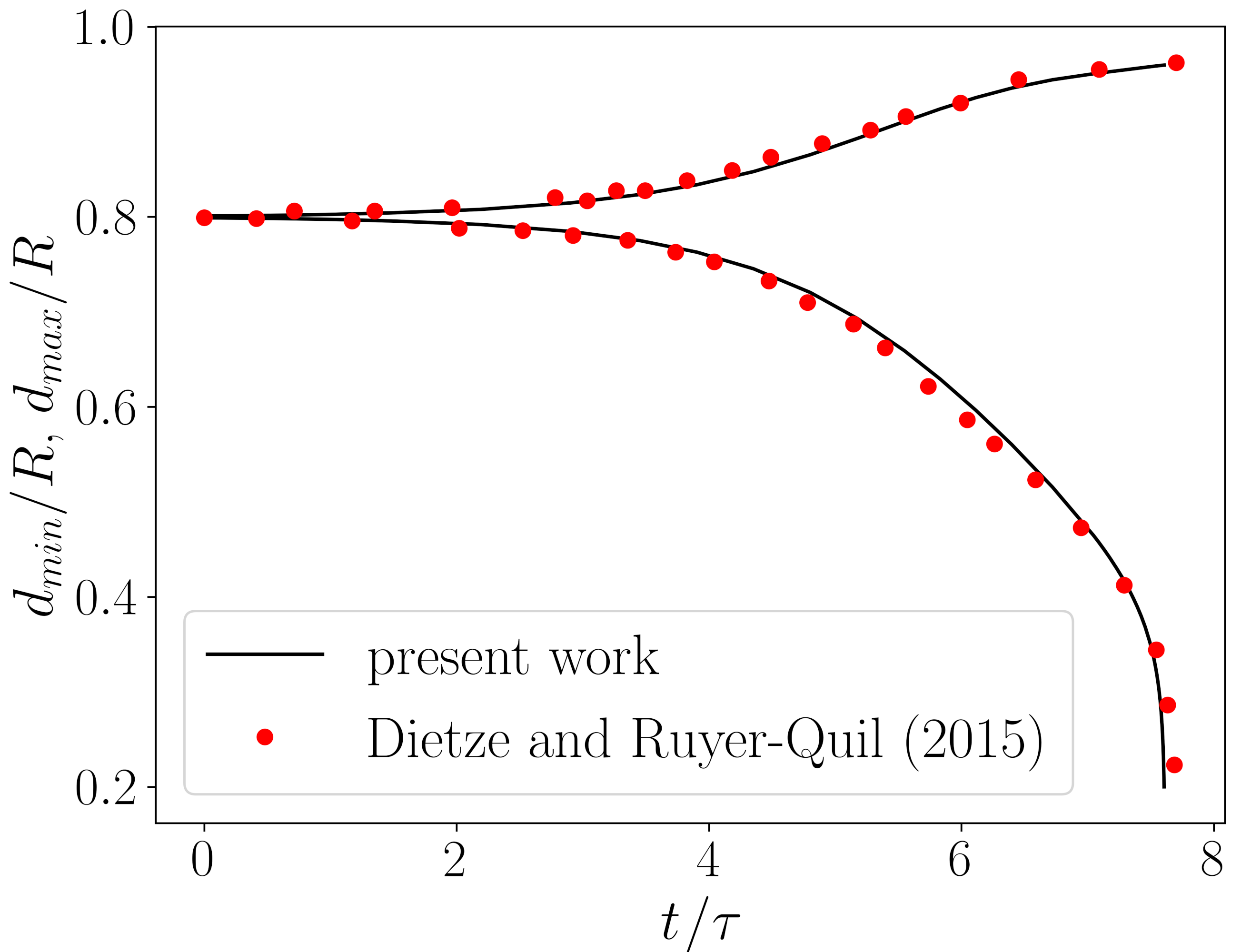}% Here is how to import EPS art
\caption{\label{fig:validation} 
Comparison of a simulation of the fully-coupled WRIBL model with the corresponding results from figure 3(\textit{d}) of \citet{Dietze2015}. The latter data (red markers) were approximated using the data extraction tool in Origin (OriginLab, USA). The evolution of the minimum and maximum radial position of the interface is compared, for a case in which the interface pinches-off. A Jupyter notebook that simulates the WRIBL model and generates this figure is available at \href{https://cocalc.com/share/public_paths/dade5dbc44da89bf9c905d19b83941294ff7914b}{https://cocalc.com/share/public{\textunderscore}paths/dade5dbc44da89bf9c905d19b83941294ff7914b}.}
\end{center}
\end{figure}

\section{Validation with \citet{Dietze2015}}\label{app:validate}

As a check on our derivation of the WRIBL equations, and to validate our numerical simulations, we have compared our results with those reported in \citet{Dietze2015} and found excellent agreement. An example is shown in figure~\ref{fig:validation}, which plots the time trace of the maximum and minimum of the interface profile for an air-mucus flow in which the interface ultimately pinches-off. Here, time is normalized by the inertialess approximation of the linear-growth timescale of the fastest-growing instability mode, which when scaled by $ R/U$ is $\tau = 6 ({\mu_m U d_0}/{\gamma})\left(\alpha^4(1/\alpha^2 - 1)(1/d_0-1)^2(1/d_0^2-1)\right)^{-1}$ where $\alpha = {2\pi d_0 R}/{\Lambda_{RP}}$. 

The \textit{Jupyter Notebook} associated with figure~\ref{fig:validation} (see the caption) demonstrates how the WRIBL equations are simulated, by first importing symbolic expressions for the coefficients from text files, then discretizing using central differences, and finally time-stepping with the LSODA algorithm using the solve{\textunderscore}ivp function in Python.

% \section{Effect of airflow on mucus when the cilia speed is low}\label{app:validate}

% \begin{figure}
% \includegraphics[width=1.0\textwidth]{image/decoupled_full_streamlines.png}% Here is how to import EPS art
% \caption{\label{fig:oneway-streamlines} 
% Detailed comparison of the air and mucus flows predicted by the fully-coupled and one-way coupled WRIBL models; the latter ignores the influence of air on the mucus film. Panels (\textit{a}) and (\textit{b}) are temporal zooms of figures~\ref{fig:oneway-vel}(\textit{b})  and \ref{fig:oneway-vel}(\textit{a}), respectively, showing the temporal variation of the mean air velocity (below the hump) and the mean mucus velocity. The line-legend is the same as that in figure~\ref{fig:oneway-vel}. Five time-points are chosen, labelled $t_1$ to $t_5$, corresponding to times when the airflow is opposite to cilia-transport ($t_1$), is in the same direction as cilia-transport ($t_2$), and is reversing ($t_3 - t_5$). The streamlines in the air and mucus at these five times are compared in panels (\textit{c}-\textit{l}) for the fully-coupled (left column, with inset time-labels) and one-way coupled (right column) models.}
% \end{figure}

\FloatBarrier

\bibliographystyle{jfm_arxiv}
\bibliography{mucus}% Produces the bibliography via BibTeX.

\end{document}